\documentclass[fleqn,10pt]{wlscirep}
\usepackage[T1]{fontenc}
\usepackage[utf8]{inputenc}
\usepackage{xcolor, authblk, csquotes, indentfirst, pdflscape, graphicx, hyperref, adjustbox, amsmath}
\usepackage[english]{babel}

\newcommand{\mbt}{$\mathrm{MnBi}_2 \mathrm{Te}_4$}
\newcommand{\mbs}{$\mathrm{MnBi}_2 \mathrm{Se}_4$}
\newcommand{\mst}{$\mathrm{MnSb}_2 \mathrm{Te}_4$}
\newcommand{\gbt}{$\mathrm{GeBi}_2 \mathrm{Te}_4$}
\newcommand{\mabt}{$\mathrm{Mn}_{1-x} A_x \mathrm{Bi}_2 \mathrm{Te}_4$}
\newcommand{\mgbt}{$\mathrm{Mn}_{1-x} \mathrm{Ge}_x \mathrm{Bi}_2 \mathrm{Te}_4$}

\newcommand{\mgpsbt}{$\mathrm{Mn}_{1-x} [\mathrm{Ge}, \mathrm{Sn}, \mathrm{Pb}]_x \mathrm{Bi}_2 \mathrm{Te}_4$}
\newcommand{\lbst}{$\mathrm{LaBi}_{1-x}\mathrm{Sb}_x\mathrm{Te}_3$}
\newcommand{\kgm}{$K\!\mathit{\varGamma}\!M$}
\newcommand{\kgz}{$K\!\mathit{\varGamma}\!Z$}
\newcommand{\zgz}{$Z\!\mathit{\varGamma}\!Z$}
\newcommand{\gz}{$\mathit{\varGamma}\!Z$}
\newcommand{\lsoc}{\lambda_\mathrm{SOC}}

\title{Phase transitions, Dirac and WSM states in \mgbt}
\author[1, *]{A.M.~Shikin}
\author[1, 2]{N.L.~Zaitsev}
\author[1]{T.P.~Estyunina}
\author[1]{D.A.~Estyunin}
\author[1]{A.G.~Rybkin}
\author[1]{D.A.~Glazkova}
\author[3]{I.I.~Klimovskikh}
\author[1]{A.V.~Eryzhenkov}
\author[1,5]{K.A.~Kokh}
\author[1,4]{V.A.~Golyashov}
\author[1,4]{O.E.~Tereshchenko}
\author[6]{S.~Ideta}
\author[6]{Y.~Miyai}
\author[7,8]{T.~Iwata}
\author[7]{T.~Kosa}
\author[7,8,9]{K.~Kuroda}
\author[6]{K.~Shimada}
\author[1]{A.V.~Tarasov}
\affil[1]{Saint Petersburg State University, St. Petersburg 198504, Russia}
\affil[2]{Institute of Molecule and Crystal Physics, Subdivision of the Ufa Federal Research Centre of the Russian Academy of Sciences, Ufa 450075, Russia}
\affil[3]{Donostia International Physics Center (DIPC), 20018 Donostia-San Sebasti\'an, Basque Country, Spain}
\affil[4]{Rzhanov Institute of Semiconductor Physics, Siberian Branch, Russian Academy of Sciences, Novosibirsk 630090, Russia}
\affil[5]{Sobolev Institute of Geology and Mineralogy, Siberian Branch, Russian Academy of Sciences, Novosibirsk 630090, Russia}
\affil[6]{Research Institute for Synchrotron Radiation Science, Hiroshima University, Hiroshima 739-0046, Japan}
\affil[7]{Graduate School of Advanced Science and Engineering, Hiroshima University, Higashi-Hiroshima 739-8526, Japan}
\affil[8]{International Institute for Sustainability with Knotted Chiral Meta Matter (WPI-SKCM$^2$), Hiroshima University, Higashi-Hiroshima 739-8526, Japan}
\affil[9]{Research Institute for Semiconductor Engineering, Hiroshima University, Higashi-Hiroshima 739-8527, Japan}
\affil[*]{ashikin@inbox.ru}


\begin{abstract}
    Using angle-resolved photoemission spectroscopy (ARPES) and density functional theory (DFT), an experimental and
    theoretical study of changes in the electronic structure (dispersion dependencies) and corresponding modification of the
    energy band gap at the Dirac point (DP) for topological insulator (TI) \mgbt{} have been carried out with gradual
    replacement of magnetic Mn atoms by non-magnetic Ge atoms when concentration of the latter was varied from 10\% to 75\%.
    It was shown that when Ge concentration increases then the bulk band gap decreases and reaches zero plateau in the
    concentration range of 45\%--60\% while non-topological surface states (TSS) are present and exhibit an energy splitting
    of 100 and 70~meV in different types of measurements. It was also shown that TSS disappear from the measured band
    dispersions at a Ge concentration of about 40\%. DFT calculations of \mgbt{} band structure were carried out to identify
    the nature of observed band dispersion features and to analyze a possibility of magnetic Weyl semimetal state formation
    in this system.  These calculations were performed for both antiferromagnetic (AFM) and ferromagnetic (FM) ordering
    types while the spin-orbit coupling (SOC) strength was varied or a strain (compression or tension) along the $c$-axis
    was applied. Calculations show that two different series of topological phase transitions (TPTs) may be implemented in
    this system depending on the magnetic ordering. At AFM ordering transition between TI and trivial insulator phase goes
    through the Dirac semimetal state, whereas for FM phase such route admits three intermediate states instead of one (TI
    --- Dirac semimetal --- Weyl semimetal --- Dirac semimetal --- trivial insulator).  Weyl points that form in FM system
    along the \gz{} direction annihilate when either the SOC strength decreases or a sufficient tensile strain is applied,
    which is accompanied by the corresponding TPTs. Model calculations of local magnetic ordering influence in AFM \mgbt{}
    was carried out by alternating Mn layers and Ge-doped layers and showed that the magnetic Weyl semimetal state in this
    system is reachable at a Ge concentration of approximately 40\% without application of any external magnetic fields.
\end{abstract}

\begin{document}
\flushbottom
\maketitle
\thispagestyle{empty}

\section*{Introduction}

Magnetic topological insulators (TIs) have attracted significant interest from researchers over the past 10 years, both
from a fundamental perspective (see, for example, \cite{1,2,3,4,5,55}) and due to their high prospects for effective use
in modern topological antiferromagnetic spintronics and magnetoelectronics (see, for example, \cite{6,7}).  Recently, a
series of intrinsic magnetically-ordered TIs from the \mbt{} family have been successfully synthesized and studied (see,
for example, \cite{7,8,9,10,11,12,13,14,15,16,17,18,19,20,21}). These materials, due to the ordering of magnetic atoms
and their increased concentrations (compared to magnetically doped TIs), are characterized by an enhanced interaction
between magnetism and topology. This leads to significantly higher temperatures for realizing the quantum anomalous Hall
effect (QAHE) and other quantum topological effects compared to magnetically doped TIs \cite{22,23,24,25,26}.  For thin
layers of \mbt{}, QAHE has been observed at a temperature of 1.4 K. This temperature increases to 6.5 K, and even up to
45 K, when transitioning from antiferromagnetic (AFM) to ferromagnetic (FM) interlayer coupling under an applied
external magnetic field \cite{22,61}.  Therefore, the search for \mbt{}-based compounds with strong spin-orbit coupling
(SOC) and FM interlayer coupling is a promising area for achieving “high-temperature” QAHE.  Additionally, this TI is
characterized by an increased energy gap at the Dirac point (DP) in topological surface states (TSS). This energy gap
can vary widely, from 70--90 meV to several meV \cite{8,9,10,11,12,13,14,15,16,17,18}, and even to gapless dispersions
\cite{27,28,29}, potentially making it possible to control the aforementioned topological quantum properties.  One
possible method to change the gap size and reach the region of topological phase transition (TPT) from the topological
to the trivial phase is the controlled replacement of heavy elements with lighter ones, such as replacing Te with Se or
Bi with Sb \cite{18,31,32,33}.  In this case, the system approaches a state close to the TPT, where an axion-like state
(with a non-zero gap at the TPT point) \cite{18} and the corresponding topological magneto-electric (ME) effect can be
realized.  The transition between the topological and trivial states can thus be controlled by an applied electric field
\cite{34JetpAxion}.

However, there is another possibility of modulating the energy gap at the DP and reaching the TPT point through the
partial replacement of Mn atoms with IV-group elements (Ge, Sn, Pb) \cite{35,36,37,38,39,40,41}.  In Refs
\cite{35,36,37}, it was shown that when Mn is replaced by Ge, Sn, or Pb atoms in the \mabt{}
($A=\mathrm{Ge},\mathrm{Sn},\mathrm{Pb}$) compounds, the bulk band gap decreases with an increase in the concentration
of the substituting atoms.  The gap decreases to practically zero at substituting element concentrations of 40--50\%,
with a possible transition from the topological state to the Weyl semimetal (WSM) state. Furthermore, there is a
potential reverse transition to a state similar to Mn-doped [Ge,Pb,Sn]$\mathrm{Bi}_2\mathrm{Te}_4$.

At the same time, for the \mgpsbt{} compounds, it was revealed that with an increase in the concentration of
non-magnetic atoms, the magnetic properties also change. The Néel temperature decreases from 24.5 K to 10--15 K when Mn
atoms are replaced by Ge, Pb, or Sn atoms at concentrations of 50--80\%, and further down to 5 K. Additionally, the
spin-flop transition threshold is reduced from 7 Tesla to 1--2 Tesla, and further to 0.5 Tesla \cite{42}.  This means
that, while the remagnetization in \mbt{} and the corresponding transition from the AFM to FM state requires the
application of a magnetic field of approximately 7 Tesla, systems with a 50\% replacement of the magnetic atom by
non-magnetic elements require only 1--1.5 Tesla. For higher concentrations, the remagnetization threshold decreases
further to 0.5 Tesla.  The study of the transition to the FM phase is very important, especially from the perspective of
transitioning to the WSM state with its unique properties. Therefore, investigating the possibility of realizing the
magnetic WSM state is a significant and challenging scientific task, particularly for TI \mabt{}.
 
It should be noted that interest in the study of Weyl semimetals (including magnetic ones) has increased significantly
in recent years. These materials are characterized by a topological invariant and the corresponding chiral charge, as
well as unique features in their electronic structure and formation conditions. This interest is due to the unique
properties of Weyl semimetals, such as the anomalous spin Hall effect, the chiral anomaly, chiral magnetic effects, and
large (negative) magnetoresistance. Additionally, Weyl semimetals are promising for the analysis of Majorana fermion
stabilization in unconventional superconductivity. They also offer the possibility of using the solid-state
black-hole-horizon analogy, which emerges at the Lifshitz transition between different WSM phases, in astrophysics (see,
for example, \cite{45,49,57,58,60} and references therein). In this context, the formed Weyl nodes (Weyl points) of
opposite chirality are singularities (sources) of Berry curvature (or monopoles of Berry flux), which can be regarded as
analogs of the magnetic monopole and antimonopole in k--space \cite{45,57,58}.

As for \mabt{}, it was shown in \cite{11,45,49,52,55} that a necessary condition for the formation of the WSM state in
compounds from the \mbt{} TI family is the presence of FM-type interaction in the system. Therefore, to analyze the
possibility of forming the WSM state (as well as a Dirac semimetal), this work presents a comparative analysis of
changes in the electronic structure of the \mgbt{} system with AFM and FM interactions. This analysis was carried out
under the gradual replacement of Mn atoms (a magnetic metal) with Ge atoms (a nonmagnetic element). In this study, Ge
atoms were selected as the substituent (from group IV elements Ge, Sn, Pb) due to their smallest atomic size in this
group. This choice was made to minimize changes in the crystal parameters of \mbt{} when replacing Mn atoms with Ge
atoms across a wide range of concentrations.

In this work, we present and analyze the results of theoretical calculations of the electronic structure of \mgbt{} with
varying Ge concentrations using the DFT method, in comparison with experimental photoemission studies. Additionally,
this work presents the results of modeling the potential implementation of a WSM (and corresponding changes in the
electronic structure) in \mgbt{} systems with AFM and FM interactions. This modeling involves modulating the SOC
strength and interatomic (interplanar) distance, assuming various influences on the analyzed system. The details of the
TPT from the TI state (characteristic of \mbt{}) to the trivial state, either through the Dirac semimetal (DSM) phase or
the DSM--WSM--DSM sequence, will be presented and analyzed. Factors influencing the implementation and specifics of such
topological phase transitions (TPTs) will also be identified. Finally, this study proposes the idea of a transition from
the AFM TI to the WSM state directly in the initial AFM phase (without external remagnetization) by introducing local
inhomogeneities in the AFM interaction through the gradual replacement of Mn atoms with Ge atoms in one Mn-layer.

\section*{Experimental Results}

\begin{figure}[ht!]
    \begin{center}
        \includegraphics[scale=0.5]{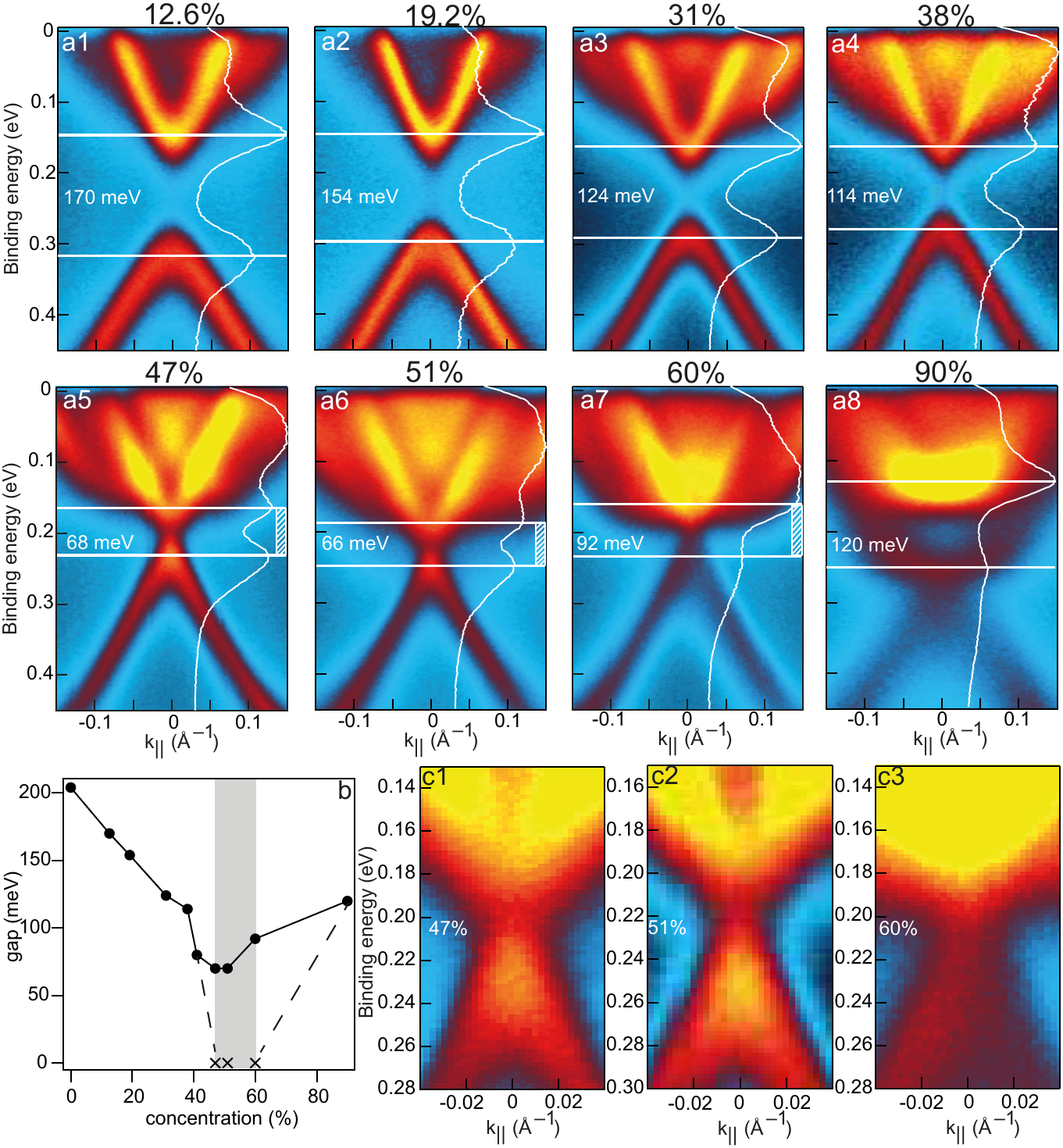}
    \end{center}
    \caption{
        (\textbf{a1}--\textbf{a8}) ARPES data for \mgbt{} with Ge concentrations ranging from 12\% to 90\%, measured
        using a He lamp ($h\nu = 21.2$~eV). White lines represent EDCs at the $\Gamma$-point. (\textbf{b}) Dependence of
        the band gap on Ge concentration, shown as the distance between peaks in the EDCs (solid line with dots) and the
        non-zero signal intensity between peaks for concentrations from 47\% to 60\% (dashed line with crosses).
        (\textbf{c1}--\textbf{c3}) Detailed view of the band dispersions near DP for Ge concentrations of 47\%, 51\%,
        and 60\%.
    }
    \label{fig1}
\end{figure}

Fig.~\ref{fig1}~(a1--a8) presents ARPES data at a photon energy of 21.2~eV for \mgbt{} with varying Ge concentrations,
ranging from 12\% to 90\% relative to the initial Mn atom concentration in \mbt{} (100\%). At this photon energy, the
ARPES data predominantly reflect bulk-derived states \cite{37,43} and non-topological surface states, providing insight
into changes in the bulk bandgap size and the nearby valence and conduction bands (VB and CB) with increasing Ge
concentrations. Notably, TSS in the bulk band gap region at the $\Gamma$-point are barely visible in these measurements
for \mbt{}.

The white lines in Fig.~\ref{fig1}~(a1--a8) represent the energy distribution curves (EDC) for the electron density
states, measured directly at the $\Gamma$-point. These curves demonstrate changes in the bulk energy gap, considering
contributions from non-topological surface states and the bulk Bi $p_z$ states at the upper edge of the VB, as detailed
in the theoretical calculations below. Fig.~\ref{fig1}~(b) shows the variation in the energy separation between the peak
maxima in the EDC around the $\Gamma$-point with increasing Ge concentration. The energy positions of the maxima on the
EDC are indicated by horizontal white lines.

As the Ge concentration increases, states contributed by Ge atoms, significantly influenced by interactions with
introduced Ge atoms, gradually emerge in the ARPES data. These states appear as "Rashba-like" states at the edges of the
cone-like dispersions. Additionally, some smeared states energetically localized within the original bulk band gap
appear in the spectra at Ge concentrations of 47--60\%. These states become more clearly observable in photoelectron
spectra under laser excitation (refer to Fig.~\ref{fig2} below).

Analysis reveals a gradual decrease in the initial bulk band gap size as Mn atoms are replaced by Ge atoms, observed up
to concentrations of 47--60\%. EDC at the $\Gamma$-point at concentrations of 47--51\% demonstrate a consistent measured
gap (peak splitting) of about 66--68~meV. Concurrently, a less pronounced cone of bulk states emerges within this gap
(energy state splitting), depicted by shaded areas in Fig.~\ref{fig1}~(a5--a7), which are more clearly visible at the
detailed representations in Fig.~\ref{fig1}~(c1,~c2,~c3). These states, along with Rashba-like states, shift towards the
initial DP, tending to form an apparent zero band gap at these concentrations. The apparent zero gap sizes at Ge
concentrations between 47 and 60~meV are marked by crosses and by grey region in Fig.~\ref{fig1}~(b).

The states exhibiting an energy splitting of about 66--68 meV (clearly discernible in the EDC) are likely formed due to
the inclusion of non-topological surface states (see the analysis of the theoretical calculations below). Moreover, the
formed Rashba-like states gradually shift towards higher binding energies with increasing Ge concentration, concurrently
hybridizing with cone-like states from the CB. As demonstrated in the theoretical calculations, this hybridization leads
to the formation and constancy of the observed 66--68~meV state splitting at Ge concentrations of 47 and 51\%. At
concentrations around 60\%, the Rashba-like states intersect with lower cone states, eventually resulting in the
formation of the apparent zero band gap. As the Ge concentration approaches 90\%, the states of the upper cone shift
back towards lower binding energies, again forming a pronounced band gap. The ARPES data then resemble those more
characteristic of Mn-doped \gbt{}, featuring the bulk band gap and TSS branches in the region of their intersection with
the edge of the VB (for comparison, see \cite{36,37} for \gbt{}).

The observed decrease in the band gap, culminating in an apparent zero gap with increasing Ge concentration, and the
formation of a plateau with a minimum band gap in the concentration range of 45--55\%, followed by a transition to the
electronic structure of Mn-doped \gbt{} (at 90\%), is consistent with previously reported data in the literature. This
pattern has been observed both in systems where Mn atoms are replaced by Ge atoms \cite{36,37}, as well as for
substitutions with other Group IV elements such as Sn \cite{35} and Pb \cite{38}. However, the underlying reason for the
formation of the minimum band gap plateau remains unclear.

To identify the contribution of the TSSs to the formed dispersions and other changes introduced by the Ge concentration
into the electronic structure near the bulk bandgap edges, detailed measurements were carried out using high energy- and
angle-resolution ARPES with laser radiation ($\mu$-ARPES and SpinLaser ARPES with $h\nu = 6.3$~eV). These measurements
focus on the TSSs in the \mbt{} family \cite{8,13,14,15,16,17,18} and are conducted at higher angular, spatial, and
energy resolutions \cite{44,Iwata2024}. The energy resolution during these measurements was 3--5~meV, and the diameter
of the laser beam was 5--10 $\mu$m.

Fig.~\ref{fig2}(a1--a7) shows the experimental dispersions measured by $\mu$-ARPES (a1--a4,a7) and SpinLaser ARPES
(a5,a6) for \mgbt{} with Ge concentrations ranging from 12\% to 90\%. In Fig.~\ref{fig2}(a1--a7), the dispersions are
shown as $N(E)$, and in Fig.~\ref{fig2}(b1--b7), as $\mathrm d^2N/\mathrm dE^2$, for better visualization of the
observed features. The presented dispersions are obtained by summing over all measured polarizations ($s$, $p$, $c_+$,
$c_-$) of the exciting laser radiation for (a1--a4,a7) and over $s$,$p$-polarizations for (a5,a6). This was done to
improve the signal-to-noise ratio, given the small differences in the measurement geometry.

Fig.~\ref{fig2}(c1--c7) shows the corresponding EDC profiles at the $\Gamma$-point, decomposed into spectral components
to highlight the TSSs and the changes in energy splitting between different states of the CB and VB with varying Ge
concentrations.

\begin{figure}[ht!]
    \begin{center}
        \includegraphics[scale=0.43]{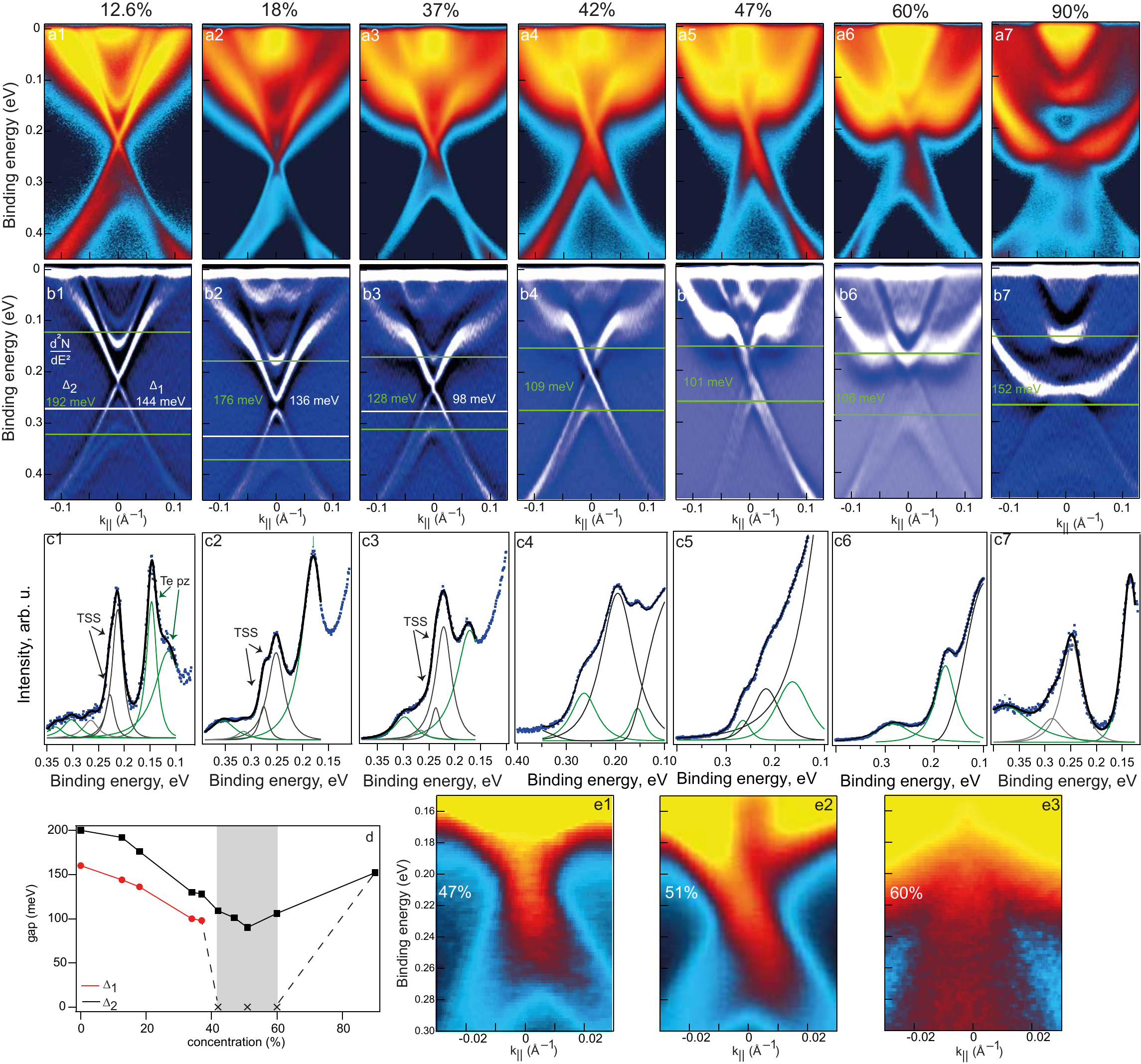}
    \end{center}
    \caption{
        Experimental dispersion profiles obtained via high energy- and angle-resolution ARPES using laser radiation ($h
        \nu = 6.3$~eV) for \mgbt{} with Ge concentrations ranging from 12\% to 90\% are presented as follows:
        (\textbf{a1}--\textbf{a7}) in $N(E)$ format, (\textbf{b1}--\textbf{b7}) in $\mathrm d^2N/\mathrm dE^2$ format,
        and (\textbf{c1}--\textbf{c7}) the corresponding density of states profiles at the $\Gamma$-point with
        decomposition into spectral components. (\textbf{d}) Dependence of the energy band gap size at the
        $\Gamma$-point on the Ge concentration. Crosses indicate the formation of an apparent zero energy bulk gap at Ge
        concentrations between 42\% and 60\%. (\textbf{e1--e3}) Detailed representation of the dispersion dependencies
        ($N(E)$) for Ge concentrations between 47\% and 60\% using $s$-polarization of laser radiation in the vicinity of
        the initial DP at the $\Gamma$-point.
    }
    \label{fig2}
\end{figure}

At low Ge concentrations, TSS branches, as well as the nearest states in the VB and CB, are clearly visible in the
presented dispersion dependencies. In Fig.~\ref{fig2}(a1), at a Ge concentration of 12\% and a temperature of 16--18~K,
the measured data shows the splitting of the Te $p_z$ states in the CB and the (Bi-Te) $p_z$ mixed states in the VB,
localized in the regions of binding energies of 0.10--0.14~eV and 0.31--0.34~eV, respectively. The energy separation
between these states in the CB and VB is indicated by horizontal green lines marked as $\Delta_2$.

According to the literature, at temperatures below the Néel temperature (24.5--25.0~K for \mbt{} \cite{8,9,13} and
25--26~K for 12\% Ge \cite{42}), the Te $p_z$ states (due to magnetic ordering in the system) are energetically split
(see Figs.~\ref{fig2}(a1,b1) for Ge 12\%). However, at temperatures above the Néel temperature, the magnitude of the
splitting of these states decreases to zero, similar to the findings in \cite{8,13,42}.

The spectra for higher Ge concentrations in Fig.~\ref{fig2} were measured at temperatures above the Néel temperatures
corresponding to these concentrations \cite{42}. Therefore, the dispersions shown in Fig.~\ref{fig2} for higher Ge
concentrations do not show the splitting of the Te $p_z$ and (Bi-Te) $p_z$ mixed states in the CB and VB (see also the
decomposition of EDC into components in Fig.~\ref{fig2}(c1--c3)).

To compare the change in separation between the Te $p_z$ states in the CB and the (Bi-Te) $p_z$ mixed states in the VB
at different temperatures, horizontal lines corresponding to $\Delta_2$ are drawn between the midpoints of the
splittings occurring for these states. The magnitude of this state separation (shown in Fig.~\ref{fig2}) was estimated
for a Ge concentration of 12\% as the average value between pairs of the top and bottom split state separations.

The measured dispersions in Fig.~\ref{fig2} reveal states near the upper part of the VB, close to the DP, at an energy
of approximately 0.28~eV. These states likely include both the Bi $p_z$ states at the VB edge and non-topological
surface states (as discussed further in the theoretical calculations). The corresponding transitions are labeled as
$\Delta_1$. It is plausible that the peaks observed in the EDC at the $\Gamma$-point in Fig.~\ref{fig1} arise from a
mixture of contributions from the (Bi-Te) $p_z$ states in the VB and these non-topological surface states, due to
limited energy and spatial resolution. Consequently, the estimated sizes of the bulk band gap from Fig.~\ref{fig1} fall
between the values of $\Delta_1$ and $\Delta_2$ depicted in Fig.~\ref{fig2}. A more detailed analysis of these
contributions will be provided later in the discussion of theoretical calculations.

Overall, the dispersions presented in Fig.~\ref{fig2} (as well as in Fig.~\ref{fig1}) show a noticeable reduction in the
state separations $\Delta_1$ and $\Delta_2$ with increasing Ge concentration. The variations in these state separations
with Ge concentration, estimated from the decomposition of EDC, are illustrated in Fig.~\ref{fig2}(d) using black and
red circles, respectively. These state separations reach average values of approximately 100~meV within the
concentration range of 42--51\%, where their constancy is observed, consistent with the trends in Fig.~\ref{fig1}.

It should be noted that the TSSs are clearly visible in the high energy- and angle-resolution ARPES data only up to a Ge
concentration of approximately 37--42\%. The energy splitting for the TSSs in the presented data is mainly in the range
of 10--15~meV. However, at a Ge concentration of 47\%, the TSSs are no longer visible in the measured data. In the
region of the DP, the contribution of the bulk Ge-contributed states is more likely to appear, leading to the formation
of an apparent zero energy gap at the $\Gamma$ point. These states are further illustrated in Fig.~\ref{fig2}(e1--e3),
showing data measured at Ge concentrations between 47\% and 60\% using s-polarization of laser radiation in the vicinity
of the initial DP at the $\Gamma$ point. Notably, no visible gap is observed in these dispersions for this range of Ge
concentrations. This suggests that at these Ge concentrations, the bulk states with the apparent zero gap at the
$\Gamma$ point are formed within the states with a peak energy splitting of about 100 meV (see corresponding EDCs in
Fig.~\ref{fig2}(c1--c7)).

At the same time, in addition to the states with an apparent zero energy gap inside the states with an energy splitting
of about 100~meV, pronounced Rashba-like states are also observed in the measured dispersions. These Rashba-like states,
as depicted in Fig.~\ref{fig1}, shift towards higher binding energies with increasing Ge concentrations, presumably
hybridizing with states from the upper CB cone in the region of the $\Gamma$-point. This hybridization leads to the
formation of peaks in the EDC profiles with a visible energy splitting of about 100~meV. Within the energy region inside
these states (peaks), as already noted, a non-zero density of electronic states is observed, resulting in the formation
of an apparent zero gap at the $\Gamma$-point.

\section*{Results of Theoretical Calculations}

\subsection*{AFM TI \mgbt. Calculations of Surface States}

\begin{figure}[ht!]
    \begin{center}
    \includegraphics[scale=0.5]{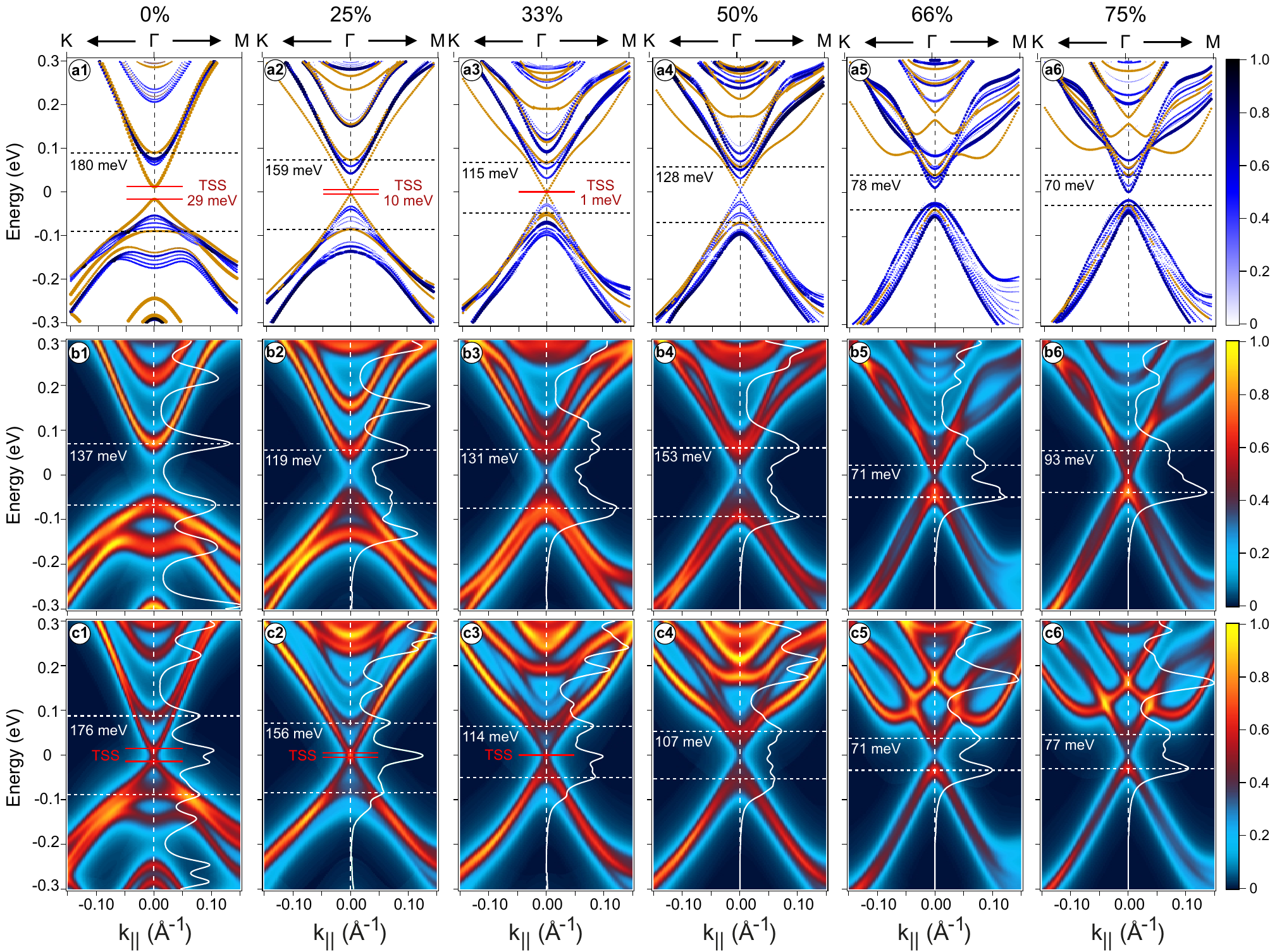}
    \end{center}  
    \caption{
        (\textbf{a1}--\textbf{a6}) Calculated band structure of a 12~SL thick slab of \mgbt{} in the \kgm{} direction in
        the BZ for Ge concentrations from 0\% to 75\%. Surface states are highlighted in light brown, and bulk states in
        black-blue-white. Horizontal dashed black lines indicate the energy positions of non-topological surface states
        at the $\Gamma$ point. Horizontal solid red lines indicate the TSS positions, which are present for Ge
        concentrations up to 33\%. (\textbf{b1}--\textbf{b6}) and (\textbf{c1}--\textbf{c6}) depict the surface and bulk
        band structures, respectively, using a Lorentzian smearing of 30~meV FWHM. Each image contains EDCs taken at the
        $\Gamma$ point (white curves). Horizontal white lines indicate the energy positions of EDC peak centers, while
        solid red horizontal lines show the TSS positions from Figs.~(a1--a6).
    }
    \label{fig3}
\end{figure}

Fig.~\ref{fig3}(a1--a6) presents the results of electronic structure calculations for a 12 septuple layer (SL) thick
\mgbt{} slab with Ge concentrations varying from 0\% to 75\%. Electronic states with more than 15\% localization in the
first SL are considered surface states, highlighted in brown, while contributions from atoms in the 3rd, 4th, and 5th
SLs are classified as bulk states, represented in black-blue. Horizontal dashed lines illustrate the energy splitting at
the $\Gamma$ point between non-topological surface states. Figs.~\ref{fig3}(b1--b6) and \ref{fig3}(c1--c6) represent the
corresponding band structures showing either bulk or surface states, respectively, using a Lorentzian broadening of
30~meV FWHM to simulate ARPES data. TSS are highlighted with red lines. White curves in the center of each image depict
EDCs of bulk and surface states taken at the $\Gamma$ point for comparison with experimental EDCs. Horizontal dashed
lines in bulk EDCs indicate peak positions, while in surface EDCs, they indicate non-TSS peak positions. Solid red lines
in surface EDCs correspond to TSSs in surface band structures.

Panel~\ref{fig3}(a) shows that the energy splitting between the VB and CB decreases as Ge concentration increases.
Gapped TSSs define the band gap at low Ge concentrations, decreasing close to zero at 33\%. Further increases lead to a
reopening of the band gap formed by bulk states, reaching up to 78~meV at $x=66$\% with a slight decrease to 70~meV at
$x=75$\%.

Simulated ARPES data in panel~\ref{fig3}(b) suggest that estimating the band gap based on EDC peaks may overestimate it.
This inconsistency can be attributed to surface sensitivity in experimental measurements. Panel~\ref{fig3}(c),
accounting for surface states, provides a better description, aligning more closely with experimental findings.

Non-topological surface states, shown in light brown, also contribute to the band dispersion of pristine \mbt{}. These
states shift towards the DP as Ge concentration increases and become prominent at 50\% concentration, exhibiting an
energy splitting of approximately 100~meV. This splitting slightly decreases to approximately 70~meV at higher
concentrations. Rashba-like surface states, which are discernible in the band dispersions, shift towards higher energies
with increasing Ge concentration. Beyond 50\% concentration, hybridization between Rashba-like states and CB states
becomes pronounced, resulting in states with an energy splitting of about 100~meV. The presence of a non-zero density of
states within the energy state splitting of approximately 70 and 100~meV at Ge concentrations of 45--60\% in the
experimental spectra suggests the formation of an apparent zero energy gap in theoretical calculations within the
concentration range of 33--50\%. The strong correlation between theoretical predictions and experimental data
underscores the validity of the presented findings.

\subsection*{AFM TI \mgbt: Calculations of Bulk Band Structure}

The bulk band structure calculations were conducted to explore the possibilities of TPTs in the \mgbt{} system. This
direction is important because the \gbt{} system (100\% substitution) exhibits band inversion at the $Z$ point, whereas
the \mbt{} system has its DP at $\Gamma$. In our study, TPTs are investigated using Te--Bi $p_z$ band diagrams, where
each point is colored red (blue) if the contribution of Te $p_z$ orbitals is greater (lesser) than that of Bi $p_z$
orbitals. These diagrams are used to detect whether band inversions occur at specific points in the Brillouin zone.

\begin{figure}[ht!]     
    \begin{center}
        \includegraphics[scale=0.5]{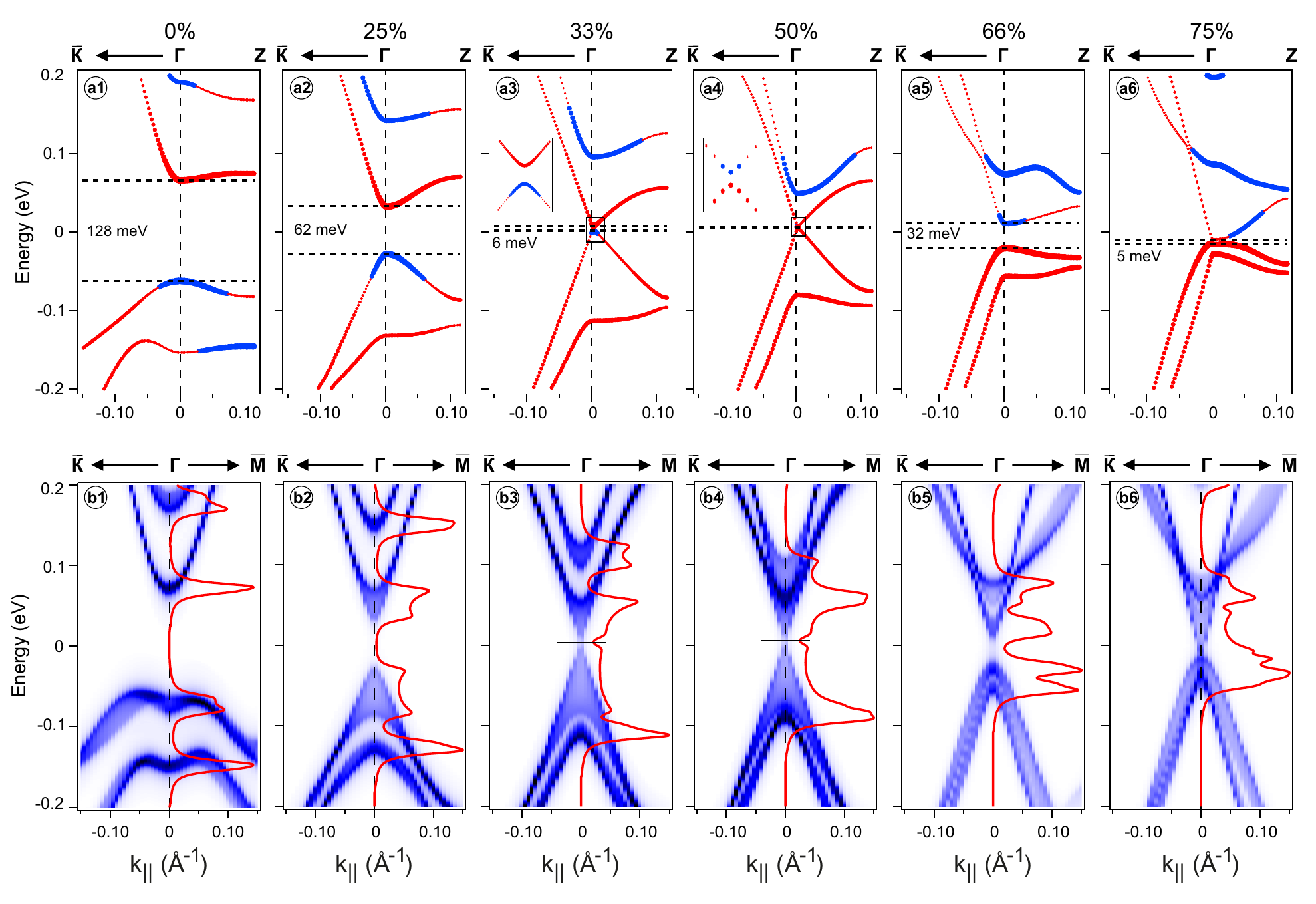}
    \end{center}
    \caption{
        (\textbf{a1}--\textbf{a6}) Bulk band structure calculations of the \mgbt{} primitive cell in the \kgz{}
        direction for different Ge concentrations up to 75\%. Blue (red) points indicate that the Bi $p_z$ orbital
        contribution at that point is greater (lesser) than the corresponding Te $p_z$ contribution.
        (\textbf{b1}--\textbf{b6}) Bulk-projected band structures along the \kgm{} direction (projection is along \gz{})
        for the same Ge concentrations. The EDC profiles at the $\Gamma$ point are shown as red curves.
    }
    \label{fig4} 
\end{figure}

Fig.~\ref{fig4}(a1--a6) shows the results of bulk band structure calculations of \mgbt{} for $x = 0$, 25\%, 33\%, 50\%,
66\%, and 75\% along the \kgz{} direction. The bulk band gap, highlighted with horizontal dashed lines, diminishes with
increasing Ge concentration. The Bi $p_z$ orbital contribution (blue points) dominates at the top of VB near the
$\Gamma$ point for concentrations up to 33\%, where the band gap reaches a value of 6~meV. This band ordering
corresponds to that of \mbt{}, indicating that the system is in the TI phase. Additional calculations for a Ge
concentration of 40\% (Fig.~{1S} Suppl. Inform.) show a band gap value of 9~meV, where Bi $p_z$ still dominates at the top
of VB.

Further increase in Ge content to 50\% leads to a band inversion within a small region around the $\Gamma$ point, while
the bulk gap is almost zero. The band ordering, where the VB is dominated by Te $p_z$ orbitals everywhere and the Bi
$p_z$ contribution dominates a small region of VB around the DP, is characteristic of a trivial insulating phase
\cite{18,30}. This suggests that the TPT into a trivial state may occur at some concentration $x_0$ slightly lower than
50\%. This \enquote{trivial} band ordering persists at concentrations of 60\% and 66\% (Fig.~1S). The bulk band gap
behaves similarly to the corresponding slab calculations in Fig.~\ref{fig3} when the Ge concentration is further
increased: it increases slightly up to 32~meV, but then decreases down to 5~meV at $x = 75\%$.

According to Fig.~{1S} Suppl. Inform., at a concentration of 80\%, the bulk band structure resembles that of the \gbt{}
state. However, the band analysis is hindered by the fact that the unit cell of \gbt{} is noticeably different from that
of \mbt{} from a geometric standpoint; for example, the cell vector length is different \cite{37}.

Fig.~\ref{fig4}(b1--b6) illustrates the bulk-projected (along \gz{}) band structure in the \kgm{} direction, allowing
for a more direct comparison with the experimental data. The resulting EDC profiles at the $\Gamma$ point are indicated
by white lines.

\subsection*{Investigation of Possible TPTs and Formation of WSM state in \mgbt{}: Model Calculations and Analysis}

While the formation of a plateau with a minimum energy gap in the extended concentration range of 30--50\% is evident,
the underlying physical processes responsible for these observed changes in the electronic structure remain to be
investigated. One potential explanation, as suggested in previous studies \cite{35,36,38}, is a possible transition to a
WSM state within the region of the minimum energy gap. In this study, we aim to conduct a detailed analysis of the
potential transition to such a state in \mgbt{} and to identify the factors influencing the occurrence of various TPTs
within this system, including the transition from the antiferromagnetic TI state to the WSM state.

It is worth noting that analyzing the possibility of forming magnetic Weyl semimetals and understanding their electronic
structure features presents a significant and challenging scientific endeavor, given the unique properties of these
materials \cite{45,57,58} and the limited depth of previous studies in the literature. Below, we will present
comparative model calculations detailing the changes in the electronic structure for the TI \mgbt, encompassing both the
AFM and FM phases, wherein states characteristic of the WSM can be explicitly discerned \cite{10,11,31,32,33,49,52}.
Additionally, we will conduct model calculations to explore the changes in electronic structure and the corresponding
TPTs possible in these systems, accounting for wide variations in the effective value of the SOC and interplanar
distances. These variations arise from the gradual replacement of magnetic Mn atoms with non-magnetic Ge atoms.

Henceforth, we will employ the term \enquote{WSM state.} Despite the band bending at the surface of the system, under
certain conditions, compensation occurs, situating the formed Weyl points near the Fermi level, similar to a typical
WSM. Notably, for theoretical calculations that do not consider this band bending, the Weyl points are already situated
near the Fermi level, i.e., within the region of zero energy values. To elucidate the factors leading to the formation
of a magnetic WSM state, we will meticulously analyze and compare the changes in electronic structure in the \gz{}
direction of the Brillouin zone. This direction is pivotal as it corresponds to the region where, according to prior
studies \cite{10,11,31,32,33,49,52}, a transition to the magnetic WSM phase should occur in \mbt{}-based systems.

Simultaneously, we postulate that at concentrations of Ge atoms ranging from 40\% to 60\%, characterized by a zero-like
band gap, local violations of the AFM order begin to emerge, or local ferrimagnetic inclusions form. These phenomena can
potentially catalyze the formation of the WSM state. At these concentrations, the energy gap at the $\Gamma$-point
closely approximates the minimum, and the system may exhibit electronic structure features indicative of the transition
to this state.

\subsection*{FM \mgbt: Dependence on the Concentration of Ge Atoms}

It is worth highlighting that one of the defining characteristics of the transition to the WSM state
\cite{45,46,47,48,49,52,57,58} is the intersection of states with different parities from the VB and CB without the
emergence of an energy gap at these state intersections. Concurrently, the transition from AFM TI to the trivial state
suggests a possible transition marked by avoided-crossing effects and the formation of a band gap at the intersection
point (for comparison, see \cite{18,30,34JetpAxion}).

In studies investigating the formation of Weyl semimetals and the intricacies of their electronic structure, as well as
the conditions conducive to the transition to the WSM state \cite{31,32,33,45,46,47,48,52,57,58}, it has been
demonstrated that such a transition may be triggered by the violation of one of two symmetries: spatial symmetry or
time-reversal symmetry, which may arise during the transition from AFM to FM. Furthermore, within the category of
ferromagnetic Weyl semimetals (e.g., \mbt, \mst, \mbs{} \cite{31,32,33,48,52}), it is precisely the violation of
time-reversal symmetry and the establishment of FM ordering that dictate the feasibility of transitioning to the WSM
state, characterized by state intersections devoid of any energy gap formation.

Simultaneously, in studies such as \cite{10,11,49,52,55}, which focus on a comparative analysis of the electronic
structure of magnetic TIs directly from the \mbt{} family in the FM phase (where time-reversal symmetry is violated), it
has been demonstrated that the transition to the WSM state (similarly observed in other Weyl semimetals
\cite{31,32,33,45,46,47,48}) occurs through the intersection of electronic states with different parities in the \gz{}
direction of the BZ. In this scenario, two Weyl points (nodes) emerge in the dispersion curves along the \zgz{}
direction (symmetrically positioned relative to the $\Gamma$-point), i.e., perpendicular to the surface plane of the
samples under examination.

Regrettably, these points cannot be directly measured via dispersions taken along the surface; however, they may
manifest in such measurements through, for instance, modulation of the gap at the DP and the density of projected bulk
states. In light of the foregoing, to validate this hypothesis, this study conducted comparative calculations of the
electronic structure (dispersion dependencies) for the FM phase of \mgbt{}—systems exhibiting FM ordering—under the
gradual replacement of Mn atoms by Ge atoms across a wide range of substitution concentrations.

\begin{figure}[ht!]      
    \begin{center}
        \includegraphics[scale=0.36]{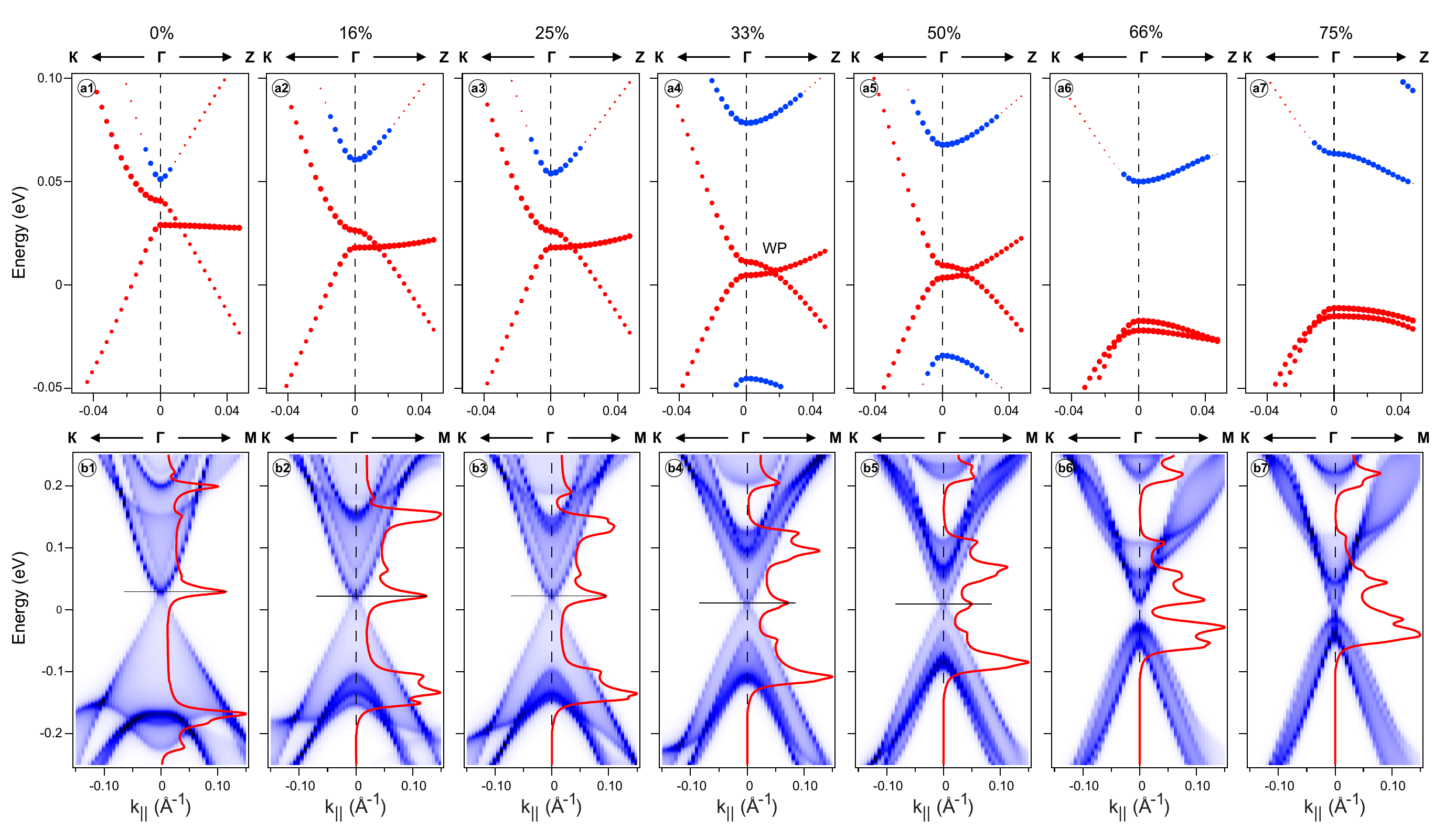}
    \end{center}
    \caption{
        (\textbf{a1}--\textbf{a7}) Theoretically calculated changes in the dispersion dependencies of bulk states in the
        \kgz{} direction of the BZ for FM TI \mgbt{} with varying Ge concentration from 0 to 75\%. Changes in the
        contributions of the Te $p_z$ and Bi $p_z$ states are shown in red and blue, respectively.
        (\textbf{b1}--\textbf{b6}) Theoretically calculated dispersion dependencies of the projected bulk states from
        the \gz{} direction to the \kgm{} directions of the BZ. The profiles of the density of electronic states at the
        $\Gamma$-point are shown by red lines, depending on the concentration of Ge atoms.
    } 
    \label{fig5} 
\end{figure}

Fig.~\ref{fig5}(a1--a7) presents the results of bulk electronic structure calculations for FM \mgbt{} at Ge
concentrations of 0, 16, 25, 33, 50, 66, and 75\%. Notably, these calculations reveal that for the FM phase of TI
\mgbt{} with Ge concentrations ranging from 0\% to 50\%, the intersection of VB and CB states characteristic of a WSM is
indeed observed in the \gz{} direction of the BZ, consistent with previous findings
\cite{10,11,31,32,33,45,46,47,48,49,52,55}. This intersection occurs at the Weyl point located at $k_\parallel =
0.015-0.02 \textup{\AA}^{-1}$ (it is marked as WP), with a similar Weyl point characterized by opposite chirality
observed in the opposite direction ($-$\gz{}), as expected for magnetic Weyl semimetals with time-reversal symmetry
violation.

Additionally, within the concentration range up to 50\% (particularly notable within the 33-50\% range), a minor energy
gap of approximately 10~meV is present at the $\Gamma$-point, exhibiting minimal variation within this concentration
range. Conversely, for Ge concentrations of 66\% or higher, the absence of state intersections in the \gz{} direction of
the BZ results in the formation of a band gap across the entire BZ, indicating a transition of the system to a
presumably trivial insulator state. In the presented dispersions, the contributions of Te $p_z$ and Bi $p_z$ states are
depicted in red and blue, respectively.

Analysis of the changes in contributions reveals that for the FM phase of \mgbt{}, the contribution of Te $p_z$ states
predominates in the region of the formed Weyl points, with no discernible inversion of contributions from Te $p_z$ and
Bi $p_z$ states observed for Ge concentrations up to 50\%. It is worth noting that the WSM state is also topological
\cite{45,57,58}. A similar trend in the contributions of Te $p_z$ and Bi $p_z$ states at the edges of branches at the
$\Gamma$-point in the vicinity of the Weyl point was also documented in \cite{31} for the FM phase of \mbs{} during the
transition to the WSM state.

With further increases in Ge concentration up to 66\%, a distinct inversion of contributions from Te $p_z$ and Bi $p_z$
states at the edges of the newly formed energy gap at the $\Gamma$-point becomes apparent. In this scenario, the
sequence of prevailing contributions from Te $p_z$ and Bi $p_z$ states at the edges of the gap more closely resembles
that characteristic of a trivial insulator.

In general, the reduction in Mn atom concentration (via their replacement with non-magnetic Ge atoms) in the FM phase
leads to a gradual decrease in the splitting between state branches at the $\Gamma$-point and a shift of Weyl points
towards the $\Gamma$-point. Between Ge concentrations of 50 and 66\%, the Weyl points annihilate at the $\Gamma$-point,
transitioning the system from a WSM to a trivial insulator state. Fig.~\ref{fig5}(b1--b7) illustrates the dispersions of
bulk states projected from the \gz{} direction of the BZ onto the \kgz{} direction. The red lines depict corresponding
EDC profiles taken directly at the $\Gamma$-point (for comparison with experiment and calculations for the AFM phase).

A key distinction between these calculations and those for the AFM phase (shown in Fig.~\ref{fig4}) is the presence of a
peak in the EDCs in the energy region of the intersection of branches at the Weyl point. This peak arises from the
contribution of states projected from the \gz{} direction and their intersection in this direction, serving as an
indicator of the realization of the WSM state with a zero band gap in the vicinity of the Weyl point. It is marked by
horizintal stripe. This peak in the EDCs at the $\Gamma$-point remains evident up to Ge concentrations of 50\%.
Conversely, for the AFM phase in the region of the DP, a dip in the EDCs was observed.

At a concentration of 66\%, coinciding with the transition to the trivial insulator state, this peak is supplanted by an
energy gap at the DP in both the \kgz{} dispersions and corresponding dispersions of bulk projected states in the \kgm{}
directions. Additionally, other peaks in the EDCs, situated further in energy from the Dirac and Weyl points, shift
towards the DP with increasing Ge concentration, mirroring experimental observations and theoretical calculations for
the AFM phase.

Interestingly, for a Ge concentration of 50\%, the energy gap between these bulk state peaks correlates even more
closely with the experiment than for the AFM phase. Moreover, if we directly compare the EDC at the $\Gamma$-point for
the AFM (see Fig.~\ref{fig4}(b4)) and FM phase (see Fig~\ref{fig5}(b5)) at the Ge concentrations in the region of 50\%
with the experimental EDC profile in Fig.~\ref{fig2}(c4,c5), then the experimental profiles (with the presence of some
peak at the initial DP instead of a dip) are more likely to correlate precisely with calculations for the FM phase with
this Ge concentration.

This may be an additional argument in favor of the possibility of a transition to the FM phase (or ferrimagnetic phase)
and the formation of the WSM state in the real \mgbt{} system at the Ge concentrations of 40--55\%, at least for the
forming local FM inclusions (see Fig.~\ref{fig8}). Now the question arises: how can a transition from the AFM TI phase
to the WSM phase actually occur under partial replacement of atoms of magnetic metals (Mn) with atoms of a non-magnetic
element (Ge) and under what conditions? If we assume that in the region of Ge concentrations of 45--55\% with a minimum
(zero-like) band gap, some features of such a transition manifest themselves in experimental spectra, then this phase
transition is plausible.

As previously noted, the WSM phase for compounds from the \mbt{} family (with the replacement, complete or partial, of
Te atoms by Se atoms or Bi atoms by Sb atoms, etc.) can be realized primarily with the FM type of interactions
\cite{31,32,33,48,49,51,52,55}. It should be noted here that the probability of the implementation of the FM phase in
the family of TIs based on \mbt{} can be greatly influenced by the formation of antisite substitution defects of the
Mn/Bi type, when Mn atoms are located at the sites of Bi atoms in the space between the magnetic layers (and vice
versa), thereby modulating the exchange interaction between magnetic Mn layers \cite{17,33,50,51}.

This leads to modulation of the magnetic interaction between the nearest magnetic layers due to a change in the
effective SOC strength since interaction between magnetic layers occurs through Mn--(Bi,~Te)--Mn superexchange between
magnetic layers through non-magnetic Bi, Te layers. The introduction of Mn atoms in place of Bi atoms can lead to the
formation of ensembles of Mn atoms with AFM (opposite) spin orientation relative to the original Mn layers, which can
locally (in the region of Mn localization at Bi sites) change the type of magnetic interaction from AFM to FM, as
observed in systems based on \mst, where the formation of antisite substitution defects of the Mn/Sb type is actually
observed \cite{50,51,52,54}.

Moreover, in systems with partial replacement of Mn atoms by Ge atoms (\mgbt{}), the formation of the above-mentioned
substitution defects of the Mn/Bi type also increases, which increases the probability of transition to the FM phase
with the growth of concentration of Ge atoms, thereby creating conditions for the transition to the WSM state. On the
other hand, because the interaction between Mn atoms in \mbt{} occurs by superexchange through Te and Bi atoms, the
replacement of Bi atoms with Ge atoms will also modulate the exchange interaction between Mn atoms, both within the
layer and between the neighboring Mn layers, stimulating the corresponding transition from AFM to the FM state, taking
into account that the formation energies of the AFM and FM phases are practically comparable for Ge concentrations from
30 to 60\%.

The transition from AFM TI to FM TI, which takes place in the $(\mathrm{MnBi}_2 \mathrm{Te}_4)_n (\mathrm{Bi}_2
\mathrm{Te}_3)_m$ family during the transition to $\mathrm{MnBi}_6 \mathrm{Te}_{10}$ compounds and further to
$\mathrm{MnBi}_8\mathrm{Te}_{13}$ (see, for instance, \cite{20,21}) can also be considered as modulation of the exchange
interaction between magnetic Mn layers through non-magnetic $\mathrm{Bi}_2 \mathrm{Te}_3$ blocks.

\subsection*{Changes in Electronic Structure Upon Modulation of SOC Strength for FM and AFM \mgbt}

To test the assumption regarding the influence of variations in the SOC strength on the modulation of Mn--(Bi, Te)--Mn
superexchange, and consequently, on the change in the magnetic interlayer Mn--Mn interaction, which determines the
possible transition between the phases of TI and WSM, as well as a trivial insulator, model calculations were conducted.
The influence of SOC strength modulation was examined for the \mgbt{} system with an initial Ge concentration of 50\% in
both the AFM and FM phases for comparison. It is presumed that for a system with 50\% Ge substitution, where the band
gap is at its minimum, the transition to the WSM phase in a real system is most probable.

\begin{figure}[ht!]
    \begin{center}
        \includegraphics[scale=0.5]{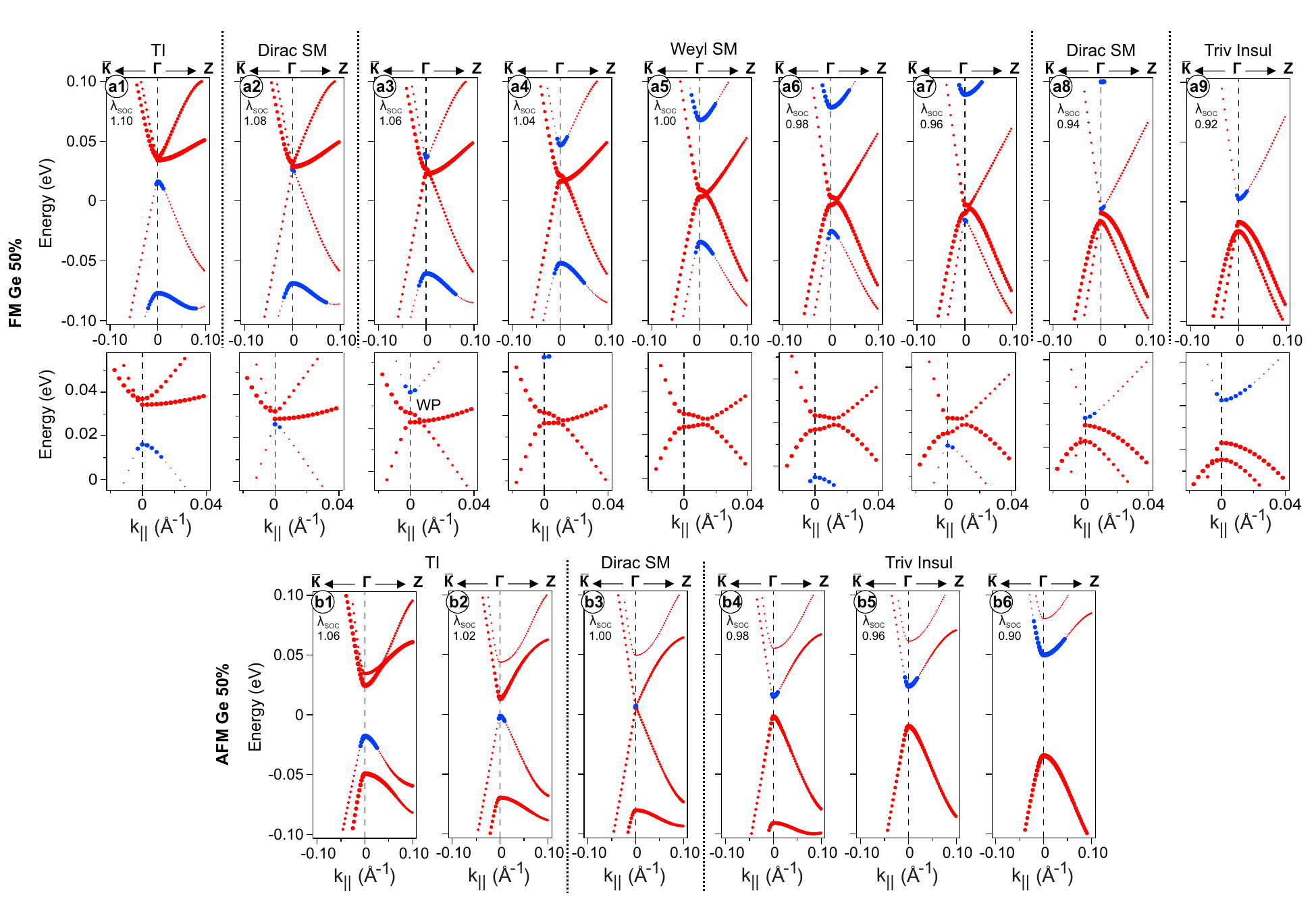}
    \end{center}
    \caption{
        Theoretically calculated changes in the dispersion dependencies of bulk states for the FM
        (\textbf{a1}--\textbf{a9}) and AFM (\textbf{c1}--\textbf{c6}) phases of \mgbt{} with $x = 0.5$ in the \kgz{}
        direction of the BZ. The modulation of the SOC strength $\lsoc$ on the Te and Bi atoms is relative to the
        initial value, taken as 1. Changes in the contributions of the Te $p_z$ and Bi $p_z$ states are shown in red and
        blue, respectively. (\textbf{b1}--\textbf{b9}) Detailed representation of dispersion dependencies in the region
        of the intersection of branches of states in the \gz{} direction of the BZ for the FM phase and the
        corresponding shift of the Weyl point with changing SOC strength.
    } 
    \label{fig6} 
\end{figure}

Fig.~\ref{fig6}(a1--a9) illustrates the theoretically calculated changes in the dispersion dependencies for the \mgbt{}
system with $x = 0.5$ and FM-type ordering (FM phase) in the \kgz{} direction in the BZ, while varying the SOC strength
$\lsoc$ relative to the initial value of 1 (corresponding to a Ge concentration of 50\%). Additionally, the
corresponding changes in the contributions of the Te $p_z$ and Bi $p_z$ states are depicted in red and blue.

The analysis of these calculations reveals a clear visibility of the intersection of state branches in the \gz{}
direction at the Weyl point (at $\lsoc = 1$), as well as a change in its position along the \gz{} direction with
variations in SOC strength. A similar Weyl point is located inversely relative to the $\Gamma$-point in the $-$\gz{}
direction.

The calculations indicate that a decrease in SOC strength (corresponding to further Mn atom replacement with Ge atoms
for Ge concentrations exceeding 50\%) results in a shift of the Weyl point in the \gz{} direction towards the
$\Gamma$-point (and the second point on the opposite side $-$\gz{}). This shift leads to the annihilation of these two
Weyl points (at $\lsoc = 0.94$) and a transition from the WSM structure to the Dirac semimetal structure with the DP at
the $\Gamma$-point. Subsequently, with further decreases in SOC strength, an absolute gap forms at the $\Gamma$-point,
which enlarges as the SOC strength diminishes.

The inversion of the contributions of the Te $p_z$ and Bi $p_z$ states at the gap edges, relative to those observed for
AFM TI (see Fig.~\ref{fig4}(a1)), suggests that a transition to the trivial insulator phase occurs as the SOC strength
decreases.

With an increase in the SOC strength relative to $\lsoc = 1$, corresponding to a decrease in the Ge concentration below
50\%, the Weyl point (and its inverse in the $-$\gz{} direction) also shifts towards the $\Gamma$-point. At $\lsoc =
1.08$, two Weyl points also annihilate at the $\Gamma$-point, resulting in the formation of the Dirac semimetal state
with a single intersection point at the $\Gamma$-point. Further increase in the SOC strength leads to the reopening of
the band gap at the $\Gamma$-point, inducing a TPT to the TI phase. In this state, the contributions of the Te $p_z$ and
Bi $p_z$ states at the gap edges follow the same sequence as for AFM TI (Fig.~\ref{fig4}(a1)), albeit in this case, it
is a FM TI. This transition from the WSM state to the TI state with increasing SOC strength is unexpected yet promising.

Thus, the SOC strength can alter the effective exchange interaction in the system, leading to a series of TPTs between
phases of magnetic topological or trivial insulators, through phases of Dirac, Weyl, and again Dirac semimetals,
depending on the increase or decrease in the SOC strength. For comparison, Fig.~\ref{fig6}(b1--b6) illustrates model
calculations of changes in dispersion dependencies in the \kgz{} direction of the BZ and corresponding changes in the
band gap size when modulating the SOC strength, relative to a system with a Ge concentration of 50\%, but with AFM
interactions. The corresponding changes in the contributions of the Te $p_z$ and Bi $p_z$ states are also shown.

The initial state (at $\lsoc = 1$) features a zero-like band gap in the region of zero energies (at the DP), closely
resembling a Dirac semimetal state. As the SOC strength decreases, the gap opens at the DP, increasing with decreasing
$\lsoc$. The inversion of the contributions of the Te $p_z$ and Bi $p_z$ states at the gap edges relative to the
sequence characteristic of AFM TIs indicates a TPT to the trivial insulator phase. Conversely, with an increase in
$\lsoc$ relative to $\lsoc = 1$, the gap at the DP reopens, with a sequence of contributions from Te $p_z$ and Bi $p_z$
states at the gap edges characteristic of AFM TI. This implies that an increase in the SOC strength leads to a
transition to the AFM TI phase, occurring within a very narrow range of changes in the SOC strength.

As a result, these calculations reveal that in the AFM phase, obtaining a plateau in the minimal band gap size
dependence with variations in the SOC strength is impossible. Moreover, for the AFM phase, intersections of states in
the \gz{} direction of the BZ are not observed, indicating that the transition to the Weyl metal state does not occur in
this phase. This transition is only possible for a system in the FM state or upon transition to the FM (or
ferrimagnetic) state.

\begin{figure}[ht!]         
    \begin{center}
        \includegraphics[scale=0.6]{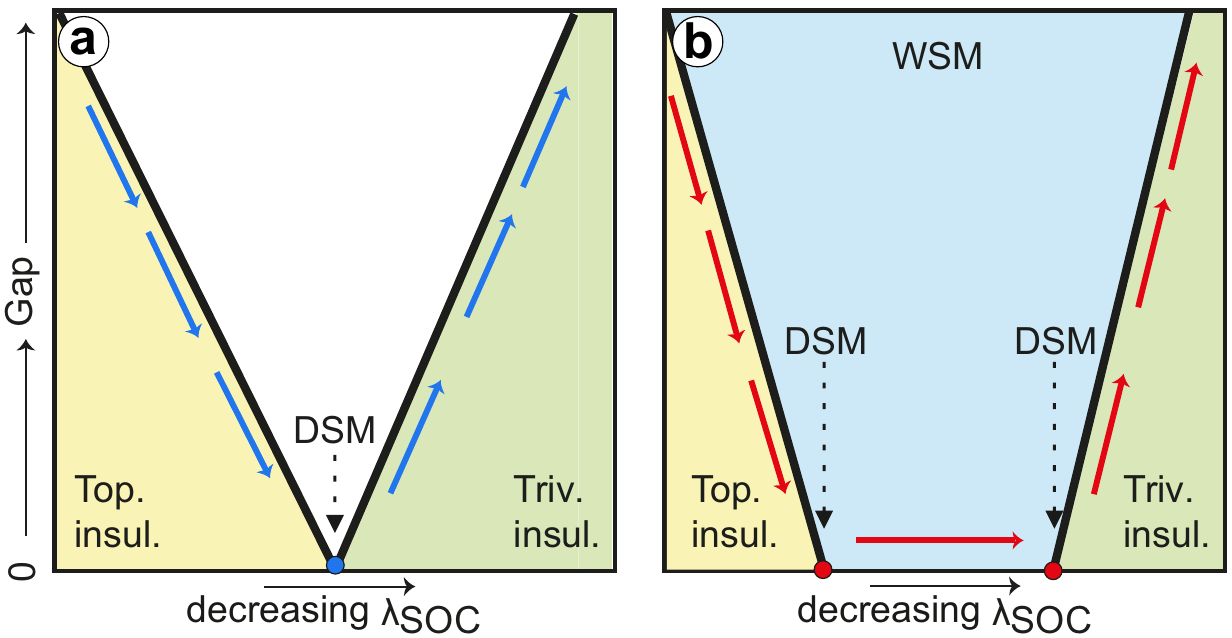}
    \end{center}
    \caption{
        Scheme of TPTs occurring in AFM TI in the absence (\textbf{a}) and presence (\textbf{b}) of an uncompensated FM
        component in the system, represented by the dependence of the gap size on the SOC strength. Changes in the
        system state on the phase diagram are shown by blue and red arrows, as well as circles.
    } 
    \label{fig9} 
\end{figure}

Schematically, the transition through the WSM state for the FM phase and its absence for the AFM phase can be
represented by the diagram shown in Fig.~\ref{fig9}. This diagram demonstrates the relationship between the modulation
of the SOC strength, the presence of an exchange field (breaking the time-reversal symmetry), and the resulting change
in the formed energy gap. For AFM phases (Fig.~\ref{fig9}~(a)), changes in the gap on the SOC strength are described by
the dependence characteristic of TPT from the state of a topological insulator to the state of a trivial insulator (see
blue arrows in Fig.~\ref{fig9}~(a)) with a minimum (ideally zero) gap at the TPT point, as in Fig.~\ref{fig6}(b).

The availability in the system of an exchange field (uniform or local one), which violates the time-reversal symmetry,
leads to the fact that the contributions to the gap formation due to the influence of the SOC and exchange interaction
contributions are compensated at a certain strength of the SOC strength. As a result, the point of the zero value of the
gap (DSM state) on the dependence of the gap on the SOC strength transforms into a line, which corresponds to the WSM
phase (but edges of this line still represent DSM states, see red lines in Fig.~\ref{fig9}~(b)). In this region, TPT
occurs from the topological to the trivial state and vice versa through the intermediate phase of the WSM.

As the exchange field decreases, the region of realization of the WSM state also decreases, shrinking to zero and
resulting in the existence of a single point of the DSM phase, as shown in Fig.~\ref{fig9}~(a). This representation
correlates with the results of works \cite{59,62,63}, showing the necessity for the formation of an intermediate phase
of the WSM under the transition from a topological to a trivial state in systems with the violation of time-reversal
symmetry (or with the violation of inversion symmetry).

\subsection*{AFM and FM \mgbt. Influence of Variation of Interplanar Distances}

Interestingly, when compared with changes in the electronic structure resulting from variations in interplanar distances
for both the AFM and FM phases (analogous to stretching and compression of the crystal lattice, akin to changes in
pressure normal to the surface), these alterations exhibit fundamental similarities to those induced by changes in SOC
strength.

\begin{figure}[ht!]
    \begin{center}
        \includegraphics[scale=0.8]{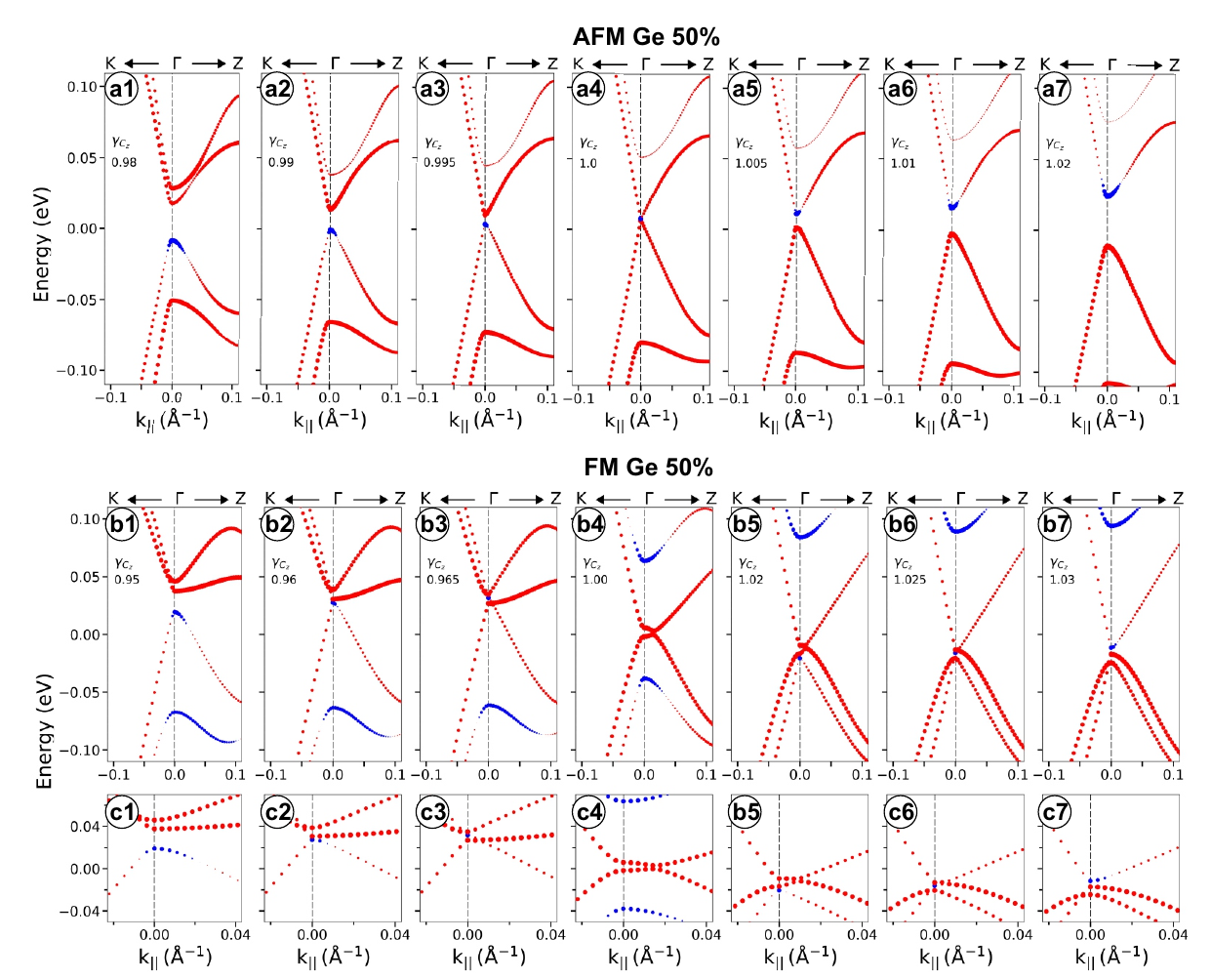}
    \end{center}
    \caption{
        Theoretically calculated changes in the dispersion dependencies of the bulk states in the \gz{} direction of the
        BZ for \mgbt{} with $x = 0.5$ for AFM - (\textbf{a1}--\textbf{a7}) and FM - (\textbf{b1}--\textbf{b7}) phases
        with variations of compression and tension of the crystal lattice parameter $c$ relative to the initial one,
        taken as $c = 1$. Changes in the contributions of the Te $p_z$ and Bi $p_z$ states are represented in red and
        blue, respectively. The insets (c1--c7) show detail in the region of the Weyl point for the FM phase (the
        intersection of the states in the \gz{} direction).
    }     
    \label{fig7} 
\end{figure}

Fig.~\ref{fig7}(a1--a7) and (b1--b7) depict the outcomes of model calculations of electronic structure changes
(dispersion dependencies) for the AFM and FM phases of the \mgbt{} system with varying interplanar distances (i.e.,
compression and tension), relative to the original system with a Ge concentration of 50\%. Here, ($+$) indicates an
increase in the interplanar distance, while ($-$) signifies its decrease. In this context, the Ge50\% unit cell
underwent compression or stretching by scaling the interplanar distance (or $z$-coordinate of the atoms) and the lattice
$c$-vector by the corresponding factor, thereby inducing compression/stretching solely along the normal to the surface
plane.

From the presented calculations (Fig.~\ref{fig7}(a1--a7)), it becomes evident that in the AFM phase, with lattice
expansion relative to the initial state ($\gamma c_{z} > 1$), a TPT to the trivial insulator state occurs. This
transition involves a reinversion of contributions from the Te $p_z$ and Bi $p_z$ states at the edges of the gap at the
DP relative to the characteristic sequence for TI, mirroring changes observed with decreasing SOC value. Conversely,
when the interplanar distance decreases relative to its initial value (i.e., when $\gamma c_{z} < 1$), transitions to
the AFM TI state occur, with a corresponding sequence of contributions from the Te $p_z$ and Bi $p_z$ states at the
edges of the resulting gap, reminiscent of TI.

These observed processes closely resemble those occurring during SOC variations, with TPTs featuring a minimum band gap
occurring at a single point. Notably, there is no plateau in the behavior of the minimum gap for the AFM phase.

For the \textbf{FM phase} (Fig.~\ref{fig7}(b1--b7)), the changes in electronic structure exhibit a series of TPTs. In
the initial state ($\gamma c_{z} = 1$) for the FM phase, the system exists as a WSM with intersecting branches of states
in the \gz{} direction of the BZ. As interplanar distances expand, the Weyl point gradually shifts towards the
$\Gamma$-point, and at $\gamma c_{z} = 1.025$, the Weyl points (nodes) annihilate at the $\Gamma$-point. Subsequently, a
TPT occurs into a state of trivial insulator. At the $\Gamma$-point, there is a reinversion of the contributions of the
Te $p_z$ and Bi $p_z$ states at the edges of the energy gap relative to the characteristic sequence of TI, mirroring
changes observed with decreasing SOC strength.

Conversely, as the interplanar distance decreases, the Weyl point also shifts towards the $\Gamma$-point, and at $\gamma
c_{z} = 0.96$, the Weyl points annihilate at the $\Gamma$-point, forming the Dirac semimetal state with a zero energy
gap at the DP. Further lattice compression at $\gamma c_{z} = 0.95$ or less leads to a transition to the TI state,
characterized by a corresponding sequence of contributions from the Te $p_z$ and Bi $p_z$ states at the edges of the
formed gap, typical of TI. Fig.~\ref{fig7}(c1--c7) shows in more detail the changes in electronic structure in the Weyl
point region.

Thus, the observed processes closely resemble those occurring under SOC strength variations. Notably, the observed
changes suggest the potential formation of a plateau with a minimum energy gap at the $\Gamma$ point in the dependence
of gap size changes with variation of interplanar distances, akin to changes induced by alterations in effective SOC
strength values.

It is worth noting that the processes of electronic structure changes in the FM phase, as derived from the calculation
results, exhibit a series of TPTs and warrant further theoretical and experimental investigation. Of course, the
possibility of transition to the FM phase remains through system remagnetization upon application of an external
magnetic field, considering that for \mgbt{} with a Ge concentration of approximately 50\%, the magnetic field magnitude
required for this remagnetization is significantly less than for pure \mbt{} \cite{42}.

\subsection*{Possibility of Transition to the WSM State Directly from the AFM TI Phase}

Analyzing the entirety of experimental and theoretical data presented above, we can infer that in the initial stage, as
the Ge concentration increases to 30\%, the system resides in the AFM TI state. This state is characterized by a
reduction in the band gap to minimum values with increasing Ge concentration, as evidenced by the presence of TSSs
visible in the experimental spectra until the Ge concentration reaches 37\%. Theoretical descriptions align well with
these observations, depicting changes in electronic structure for AFM TIs within these limits.

As the Ge concentration further increases, states near the DP transform into bulk states, leading to the disappearance
of TSSs from the spectra and the emergence of bulk Ge-contributed states around the DP region. Concurrently, the
accumulation of antisite defects occurs in the system, facilitating the transition of either the entire system or a
portion of it to the FM (or ferrimagnetic) state. This potentially enables local implementation of the WSM state within
Ge concentrations ranging from 40\% to 50--60\%. Moreover, electronic structure alterations, including variations in
bulk band gap size, characteristic of the FM phase, are discernible within these concentrations. However, part of the
system remains in the AFM state, manifesting corresponding electronic structure changes typical of the AFM phase.

As the Ge concentration exceeds 60--65\%, a transition to the trivial insulator phase occurs for both the FM and AFM
phases, according to calculation results. Subsequently, with further increases in Ge concentration, a shift occurs
towards Mn-doped \gbt{}. Evidence supporting the transition from an AFM TI to a trivial insulator phase with changing Ge
concentration possibly occurring through the WSM phase is provided by work \cite{59}, demonstrating that the WSM phase
serves as an intermediate phase during the transition from TI to trivial insulator states for compounds such as \lbst,
among others. This transition includes the formation of a plateau exhibiting a zero band gap in the region of such a
phase transition.

Additionally, the realization of the critical WSM phase (i.e., the plateau region with zero band gap) can be highly
sensitive to the chosen path in the parameter space, suggesting flexibility in engineering the WSM phase. Similarly, in
\cite{62}, it was shown that the phase diagram of a multilayer structure composed of identical thin films of a
magnetically doped 3D topological insulator, separated by ordinary-insulator spacer layers, contains a WSM phase. This
phase acts as an intermediate phase between topological and ordinary insulators for systems with broken time-reversal
symmetry.

In the case of a system with FM ordering, this transition through the WSM phase occurs naturally due to the breaking of
time-reversal symmetry, with the corresponding parameters being changes in SOC strength or interplanar distance in the
crystal lattice. Conversely, in the AFM phase, where changes in these parameters are insufficient, additional parameters
such as introducing an asymmetric distribution of sites for Mn atom substitution with Ge atoms or the presence of
corresponding Bi/Mn antisite defects may facilitate the transition to the WSM phase. Hence, calculations solely
involving changes in SOC strength or lattice contraction/expansion for the AFM phase do not demonstrate the formation of
the intermediate WSM phase. However, calculations depicting changes in electronic structure with varying Ge
concentration do show the presence of a plateau in the band gap minimum characteristic of the WSM.

\begin{figure}[ht!]    
    \begin{center}
        \includegraphics[width=\linewidth]{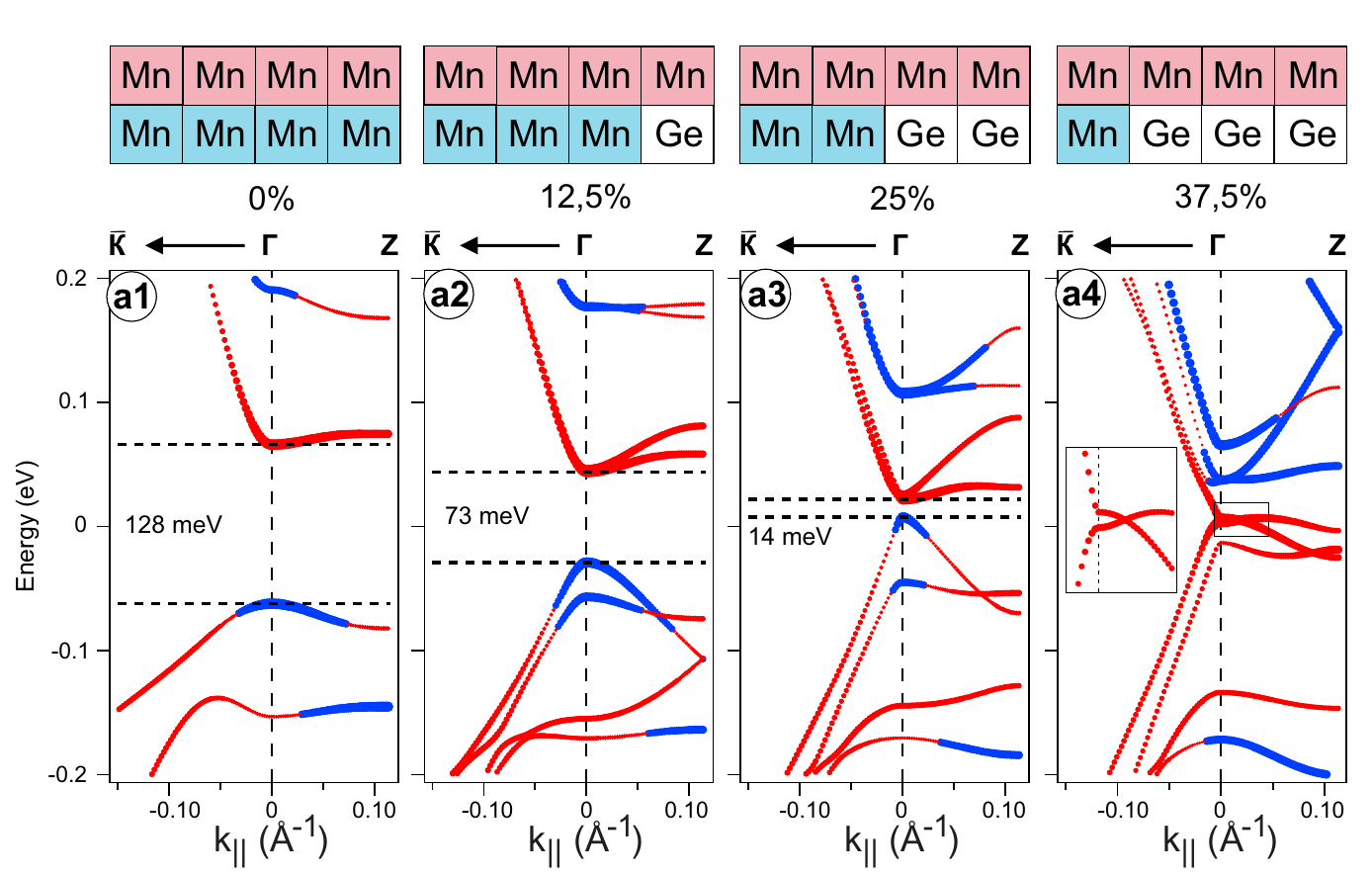}     
    \end{center}
    \caption{
        Theoretically calculated changes in the electronic structure of AFM TI \mgbt{} with varying Ge concentration
        under the condition of alternate replacement of Mn atoms with Ge through the Mn layer. Model depictions of the
        placement of Ge and Mn atoms are shown in the insets above (red and blue colors denote AFM ordering of Mn
        layers). For the Ge atom concentration of 38\%, a more detailed representation is shown in the region of the
        intersection of states in the \gz{} direction of the BZ, characteristic of a WSM.
    }     
    \label{fig8} 
\end{figure}

An additional factor that may stimulate the transition to the WSM phase (at least locally) at certain Ge concentrations
is the asymmetry of the local magnetic order when Mn atoms are replaced by Ge atoms. Fig.~\ref{fig8}(a1--a4) presents
the results of model calculations depicting changes in dispersions for \mgbt{} in the \kgz{} direction of the BZ with a
change in Ge concentration under the condition of alternating replacement of Mn atoms with Ge, not in each Mn layer, but
through the layer (in contrast to disordered placement of Ge atoms).

Model depictions of Ge and Mn atom placement for such a system are also provided in the insets above. This structure
bears some resemblance to the multilayer structure composed of a stack of thin layers of magnetically doped 3D TI,
separated by insulating spaces studied in \cite{62}, where the 3D WSM phase was realized.

The presented dispersion dependencies reveal a monotonic decrease in the bulk band gap size to 14~meV with an increase
in the Ge concentration, consistent with both theoretical and experimental spectra within the concentration range of 0
to 25\%. Additionally, some local splitting of the dispersion branches can be observed, possibly associated with an
increasing disruption in the AFM interaction.

At a Ge concentration of 38\% (Fig.~\ref{fig8}(a4)), the intersection of state branches in the \gz{} direction,
characteristic of the transition to the WSM state, is observed in the vicinity of the initial DP (zero energies). This
results in the formation of corresponding Weyl point(s) in the \gz{} and $-$\gz{} directions, similar to observations in
the FM phase. This phenomenon is more clearly illustrated in the inset of Fig.~\ref{fig8}(a4).

The splitting between states adjacent to the band gap minimum at this concentration closely aligns with the
experimentally observed splitting of corresponding states in Figs.\ref{fig1} and \ref{fig2}. The distribution of Te
$p_z$ and Bi $p_z$ contributions to the states at the edges of the bulk gap behaves as expected during the transition
from an AFM TI (Figs.\ref{fig4} and \ref{fig6}) to the WSM (Figs.~\ref{fig5} and \ref{fig6}).

These calculations suggest the potential formation of the WSM state in the initial AFM TI \mgbt{} at Ge concentrations
ranging from 40\% to 55\% within the plateau region for the minimum of the bulk band gap. This occurs alongside local
violations of the AFM-type interaction, where the replacement of Mn atoms with Ge is not disordered in each Mn layer but
occurs through the Mn layer.

\section*{Details of DFT Calculations}

First-principles calculations in the framework of the density functional theory (DFT) were performed at the Computing
Center of SPbU Research park. The electronic structure supercell calculations with impurities were performed using the
OpenMX software code, version 3.9.9, which provides a fully relativistic DFT implementation with localized pseudoatomic
orbitals \cite{64,65,66} and norm-conserving pseudopotentials~\cite{67}. The exchange correlation energy in the PBE
version of the generalized gradient approximation was employed~\cite{68}.

Impurity concentrations of 25\%, 50\% and 75\% were modeled using the $2 \times 2$ supercell, which provides 4 sites for
impurity atoms; concentrations of 33\% and 66\% were obtained using the $3 \times 1$ supercell, which allows 3 sites for
impurity atoms; impurity concentration of 16\% was achieved in the $6 \times 1$ supercell, which admits 6 impurity
sites.

The surface calculations were performed using 12 SL slabs of \mbt{} with 20~\AA{} vacuum separation between crystal
surfaces. The accuracy of the real-space numerical integration was specified by the cutoff energy of 300~Ry, and the
total energy convergence criterion was $10^{-6}$~Hartree. Uniform reciprocal space meshes for Brillouin zone sampling
were specified as $7 \times 7 \times 7$ for pristine bulk calculations, $5 \times 5 \times 1$ mesh for pristine \mbt{}
slab calculations, and $3 \times 3 \times 7$ for bulk and $3 \times 3 \times 1$ for slab supercells with impurities,
respectively.

All calculations were performed using pseudoatomic basis sets of Bi8.0-s3p2d2f1, Te7.0-s3p2d2f1, Mn6.0-s3p2d1,
Ge7.0-s3p2d2, where the pseudopotential cutoff radius is followed by a basis set specification. The Mn~$3d$ states were
treated within the DFT+$U$ approach~\cite{69} using the Dudarev scheme~\cite{70} with $U = 5.4$~eV. The resulting
magnetic moment on Mn atoms in all self-consistent calculations was equal to 5.0~$\mu_{\mathrm B}$.

\section*{Experiment Details}

The \mgbt{} single crystals were grown by the Bridgman method where $x$ is the molar ratio of Ge atoms. These crystals
are solid solutions of the compounds \mbt{} and \gbt{}, both sharing the rhombohedral crystal structure with the
$R\bar{3}m$ symmetry space group. Clean surfaces of the samples were obtained by cleavage under ultrahigh vacuum
conditions along the easy cleavage plane (0001).

Angle-resolved photoemission spectroscopy (ARPES) measurements were performed at the Rzhanov Institute of Semiconductor
Physics SB RAS (Novosibirsk, Russia) using the SPECS ProvenX-ARPES facility equipped with the SPECS ASTRAIOS 190
analyser. He I$\alpha$ ($h\nu = 21.2$~eV) radiation was used for these ARPES measurements, and Al K$\alpha$ ($h\nu =
1486.7$~eV) radiation was used for stoichiometry estimation of the samples by means of X-ray photoelectron spectroscopy
(XPS).

Laser-based micro-focus angle resolved photoemission ($\mu$-ARPES) experiments were performed at the $\mu$-Laser
(Scienta R4000 analyzer) and LaserSpin (Scienta DA30 analyzer) ARPES stations at HiSOR (Hiroshima,
Japan)\cite{44,Iwata2024}. A laser beam with photon energy of $h\nu = 6.3$~eV and photon flux of up to $10^{14}$
photons/s was used for $\mu$-ARPES measurements. The incident photon beam spot diameter was estimated to be
5--10~$\mu$m.

During the measurement process, the temperature of the samples was maintained below 30~K. The base pressure in the
analytic vacuum chamber was less than $10^{-10}$~Torr for all performed measurements.

\section*{Conclusion}

Our comprehensive study of \mgbt{} single crystals with varying Ge concentrations has provided significant insights into
the evolution of their electronic structure. Through detailed ARPES measurements, we have observed a clear correlation
between the Ge concentration and the bulk band gap behavior. Specifically, as the Ge concentration $x$ increases up to
approximately 40\%, the bulk band gap decreases to zero, indicating a TPT.

Within this concentration range, the TSS are well visible, but their intensity gradually decreases with increasing Ge
concentration; they disappear completely as the Ge content continues to increase. The band gap is assumed to be zero for
concentrations between 45--60\% due to the presence of the ARPES signal in the DP region from the bulk Ge-contributed
states. Further increases in Ge concentration result in a corresponding increase in the band gap, as the \mgbt{} system
with high $x$ behaves more like Mn-doped \gbt{}, where the band gap is finite. These bulk Ge-contributed states also
manifest as Rashba-like bands that lower their energy with increasing Ge concentration. Finally, non-topological surface
states, exhibiting an energy splitting of 70--100~meV (measured by different methods), remain constant for Ge
concentrations of 45--60\%.

DFT electronic structure calculations of 12 SL-thick slabs and bulk primitive cells of \mgbt{} with varying Ge
concentrations also show that the bulk band gap decreases as the Ge concentration increases and reaches zero values when
Ge concentrations are in the 33--50\% range. With further increases in Ge concentration above 60\%, the bulk gap
reopens. Bulk-projected band calculations also confirm the formation of a zero-gap plateau. Furthermore, DFT calculation
results confirm that the band dispersions display the presence of Rashba-like surface states in addition to TSS, which
shift towards higher binding energies with increasing Ge concentration. These states hybridize with conduction bands and
exhibit energy splitting of approximately 100 and 70~meV, which remains stable in the concentration range of 45--60\%.

To analyze the possibility and reasons for the formation of a plateau with a minimum band gap, as well as to explore the
potential for forming the state of a magnetic WSM in this system, comparative calculations of changes in the electronic
structure in the AFM and FM phases were carried out. These calculations varied both the concentration of Ge and the
modulation of the SOC strength and interplanar distance (through expansion and contraction of the system). The results
showed that for the FM phase, the TPT from the TI state to the TrI state does not occur directly. Instead, it proceeds
through a sequence of topological transformations: from the TI phase through the DSM state to the WSM phase (with the
intersection of states in the \gz{} direction) and then to the TrI phase, again through the DSM state
(TI–-DSM--WSM--DSM--TrI). This sequence of TPTs suggests the possibility of forming a plateau in the dependence of the
band gap size, as observed in the experiment. In contrast, for the AFM phase, calculations of the dependence of the
electronic structure on the SOC strength modulation and interplanar distance do not show the presence of such a plateau
with a minimum band gap. Instead, they show only the TPT from the AFM TI to the TrI state through the DSM phase in a
single-point transition.

Based on these calculations, we assumed that the presence of the plateau in the experimental dependence of the band gap
in the Ge concentration range of 35--50\% can be determined by the local transition of part of the sample to the WSM
state (presumably with local ferrimagnetic ordering), which ensures the formation of a plateau with an apparent zero
bulk gap in the region of these concentrations. To test this assumption, model calculations were carried out, which
confirmed the possibility of a transition to the state of a magnetic WSM directly from an AFM TI under the condition of
local disruption of magnetic ordering by the replacement of Mn atoms by Ge not in every Mn layer, but through the layer.
This condition creates the maximum possible local violation of the AFM-type interaction. In this case, the electronic
structure acquires the features characteristic of the state of a magnetic WSM without the need for external
remagnetization of the system.

\section*{Acknowledgements}

This work was supported by the Russian Science Foundation grant No. 23-12-00016 and the St.~Petersburg State University
grant No. 95442847. HiSOR ARPES measurements were performed under Proposals No. 23AG008, 23AU003, 23AU012, 23BU003. We
are grateful to the N-BARD, Hiroshima University for liquid He supplies. The \mgbt{} single crystals were grown using
the Bridgeman method under state assignment of Sobolev Institute of Geology and Mineralogy SB RAS \#122041400031-2.  The
project that gave rise to these results received the support of a fellowship from \enquote{la Caixa} Foundation (ID
100010434). O.E.T. and V.A.G. acknowledge  RSF No. 22-12-20024 (p-9) in the part of the XPS calibration of stoichiometry
for ARPES measurements with using He lamp at ISP SB RAS. The calculations were partially performed using the equipment
of the Joint Supercomputer Center of the Russian Academy of Sciences (https://rscgroup.ru/en/project/jscc). The authors
acknowledge SPbU Research Park \enquote{Physical Methods of Surface Investigation} and \enquote{Nanotechnology} where
the elemental composition of studied samples was investigated.

\section*{Author contributions statement}

The manuscript was written and edited by A.M.~Shikin, A.V.~Tarasov, and A.V.~Eryzhenkov. All co-authors took part in the
discussion and analysis of the experimental results. Preparation of the manuscript for publication, including the
presentation of the figures, was carried out by A.V.~Tarasov and T.P.~Estyunina. ARPES measurements with He I$\alpha$
radiation were conducted by A.M.~Shikin, D.A.~Estyunin, T.P.~Estyunina, V.A.~Golyashov and O.E.~Tereshchenko. ARPES
measurements with with laser radiation ($\mu$-ARPES and SpinLaser ARPES) were conducted by A.M.~Shikin, D.A.~Estyunin,
T.P.~Estyunina, A.G.~Rybkin, I.I.~Klimovskikh, S.~Ideta, Y.~Miyai, T.~Iwata, T.~Kosa, K.~Kuroda and K.~Shimada.
Experimental data processing was handled by D.A.~Glazkova, T.P.~Estyunina and D.A.~Estyunin. The band structure
calculations were carried out by A.V.~Tarasov, N.L.~Zaitsev, and T.P.~Estyunina. Crystals were grown by K.A.~Kokh and
O.E.~Tereshchenko. All authors reviewed the manuscript. The project was planned and supervised by A.M.~Shikin.

\section*{Competing Interests}

The authors declare no competing interests.

\section*{Data Availability}

All data generated or analysed during this study are included in this published article (and its Supplementary
Information files).

\section*{Additional information}

\textbf{Supplementary Information} The online version contains supplementary material.

\bibliography{main.bib}

\end{document}